\documentclass[3p,preprint,hyperref]{elsarticle}
\input{FFPR_aux}

\begin{document}

\begin{frontmatter}
\title{Feynman{-}Fox integrals in the physical region}

\author[torino]{Giampiero Passarino}
\ead{giampiero@to.infn.it}

\address[torino]{\csuma}


\begin{abstract}
 \noindent
Feynman integrals in the physical region are transformed into Fox functions with a special emphasys
to their cut structure.
%
\end{abstract}

\end{frontmatter}

\tableofcontents

\newpage
%
%
%

\section{Introduction \label{intro}}
The connection between Feynman integrals and Mellin{-}Barnes (MB) integrals has a long
history~\cite{oFox,compH,HS,Hus,BSid,More,Passarino:2024ugq,Dubovyk:2022obc,Kalmykov:2012rr} 
and has been systematized in \Bref{Passarino:2024ugq} where the MB integrals are related to generalized Fox 
functions~\cite{oFox,compH,BRaa,SGG,HS,Kershaw:1973km} ($\mrH$ functions).
Feynman integrals are indispensable for precision calculations in quantum field theory; generically they
are characterized by a branch cut structure determined by the Landau 
equations~\cite{Landau:1959fi,Eden:1966dnq,Nakanishi:1968hu,doi:10.1143/PTP.22.128}. 
Landau equations for a given Feynman integral~\cite{Passarino:2018wix} are a set of kinematic constraints 
that are necessary for the appearance
of a pole or branch point in the integrated function (as a function of external kinematics and masses). Landau equations
admit many families of solutions which are naturally classified as leading
Landau singularities, sub-leading Landau singularities, sub-sub-leading \etc
To give an example~\cite{Kershaw:1971rc}, in the most general one{-}loop triangle the physical Landau curve has 
six branches; when we consider the most general one{-}loop box we get $14$ branches.
A two{-}loop example~\cite{Passarino:2001jd}, is given by the following set of propagators ($\mrp$ is the external
momentum):
\bq
\mrq_1^2 + \mrm_1^2 \spc \quad
(\mrq_1 - \mrq_2)^2 + \mrm_2^2 \spc \quad
\mrq_2^2 + \mrm_3^2 \spc \quad
(\mrq_2 + \mrp)^2 + \mrm_4^2 \spp
\eq
Considering the corresponding Landau equations we obtain the leading Landau singularity (the so{-}called
anomalous threshold) for
\bq
\mrs = (\mrm_1 + \mrm_2 \pm\mrm_4)^2 \spc \qquad
\mrm_3^2 = ( \mrm_1 + \mrm_2)^2 \spc
\eq
where $\mrs = - \mrp^2$. Sub{-}leading singularities are given by $\mrs = (\mrm_1 + \mrm_2 \pm\mrm_4)^2$ 
corresponding to the normal and pseudo{-}thresholds of the
reduced diagram where the line corresponding to the propagator
$\mrq_2^2 + \mrm_3^2$ is contracted to a point.

In this work we consider the connection between Feynman integrals (defined in the physical region) and
Fox functions 
with a special attention to their behavior below and above normal thresholds.

In this paper ``physical region'' is identified with
the phase space for the corresponding process, \ie the physical region of a given process is the set of all real 
initial and final energy-momenta variables subject to the mass-shell conditions and to energy-momentum conservation. 
Solutions that correspond to points outside the physical region are on the wrong sheet.
Any process $\mrn \to \mrm$ is described by $3\,(\mrm + \mrn) - 10$ Mandelstam invariants and the physical region
is dictated by the corresponding phase space. For instance, the decay process 
\bq
\mrA(\mrQ) \to \mra_1(\mrp_1) + \mra_2(\mrp_2) + \mra_3(\mrp_3) \spc
\eq
is described by two invariants,
\bq
\mrs = - \lpar \mrQ - \mrp_1 \rpar^2 \spc \qquad
\mru = - \lpar \mrQ - \mrp_2 \rpar^2 \spc
\eq
and the physical region~\cite{Kumar:1970cr} is given by
\bq
\lpar \mrm_2 + \mrm_3 \rpar^2 \le \mrs \le \lpar \mrM - \mrm_1 \rpar^2 \spc \qquad
\mru_{-} \le \mru \le \mru_{+} \spc
\eq
\bq 
\mru_{\pm} = \mrM^2 + \mrm_2^2 - \frac{1}{2\,\mrs}\,\lpar \mrs + \mrm_2^2 - \mrm_3^2 \rpar\,
\lpar \mrs + \mrM^2 - \mrm_1^2 \rpar \pm \frac{1}{2\,\mrs}\,
\lambda^{1/2}\lpar \mrs\,,\,\mrm_2^2\,,\,\mrm_3^2 \rpar\,
\lambda^{1/2}\lpar \mrs\,,\,\mrM^2\,,\,\mrm_1^2 \rpar \spc
\eq
where $\mrM$ is the invariant mass of the incoming system and $\lambda$ is the K\"allen function.
A complete set of results can be found in \Bref{Kumar:1970cr} where
integrands containing arbitrary invariant functions of momenta are
transformed into integrals over Mandelstam{-}like variables.

General aspects of the problem are introduced and discussed in \autoref{gasp}; 
in particular we want to associate a Fox
function to Feynman integrals in such a way that the function reproduces the
correct imaginary part below and above any singularity of the
Feynman integral (inside the physical phase space).
Several examples are presented in \autoref{exa}, including virtual and real radiative corrections.
\section{General aspects \label{gasp}}
We are mostly interested in two{-}loop Feynman integrals defined as follows:
\bq
\mrI = \frac{\Gamma(\upnu)}{\sum_{i=1,\mrL}\,\Gamma(\upnu_\mrj)}\,
\int d^{\mrL}\alpha\,\delta\lpar 1 - \sum_{i=1,\mrL}\,\alpha_\mrj \rpar\,
\prod_{i=1,\mrL}\,\alpha^{\upnu_\mrj - 1}_\mrj\,
\int\,\mrd^{\mrd} \mrq_1\,\mrd^{\mrd} \mrq_2\,\mathcal{I}^{-\upnu} \spc
\eq
where $\mrL$ is the number of internal lines and propagators have non{-}canonical powers $\upmu_\mrj$ with 
$\upnu = \sum_{i=1}^{\mrL}\,\upnu_\mrj$. 
Integrating over momenta gives
\bq
\int\,\mrd^{\mrd} \mrq_1\,\mrd^{\mrd} \mrq_2\,\mathcal{I}^{-\upnu} = 
- \pi^{\mrd}\,\frac{\Gamma(\upnu - \mrd)}{\Gamma(\upnu)}\,
\mrS^{\upnu - 3/2\,\mrd}_1\,\mrS^{\mrd - \upnu}_2 \spc
\label{pfac}
\eq
where $\mrS_1$ and $\mrS_2$ are Symanzik polynomials~\cite{Weinzierl:2013yn,Bogner:2010kv}.
A few remarks on the two Symanzik polynomials are in order: both polynomials
are homogeneous in the Feynman parameters; for $l$ loops $\mrS_1$ is of degree $l$,
$\mrS_2$ is of degree $l+1$. The polynomial $\mrS_1$ is linear in each Feynman parameter. 
If all internal masses are zero, then also $\mrS_2$ is linear in each Feynman parameter. 
In expanded form each monomial of $\mrS_1$ has coefficient $+\,1$. 

Partial quadratization~\cite{Passarino:2024ugq} is a change of variables $\{\alpha\} \to \{\mrx\}\,,\{\rho\}$ such that
$\mrS_1$ depends only on $\{\rho\}$ and $\mrS_2$ is a multivariate quadratic form in $\{\mrx\}$ with
coefficients which are $\{\rho\}\,${-}dependent. Our goal is to show that Feynman integrals can be written as
combinations of Fox functions (defined on different contours) respecting the cut structure of the original integral.
Furthermore, our goal will be to define MB integrals in a way that they can be computed numerically instead
of computing multidimensional residues~\cite{Catt,Tsi,ZT}.
The numerical computation of Fox functions will be the argument of a forthcoming paper~\cite{NCFF}. 

There are three crucial points in writing a Feynman integral in terms of Fox functions:
\begin{enumerate}

\item in general $( \mra\,\mrb )^{\alpha} \not= \mra^{\alpha}\,\mrb^{\alpha}$ when $\mra, \mrb, \alpha \in\Cf$ \spc

\item the so{-}called Mellin{-}Barnes splitting~\cite{Freitas:2010nx} requires special conditions,

\item the Mellin{-}Barnes representation of a Lauricella function~\cite{FDMB,CarlsonR,Asur} requires special conditions.

\end{enumerate}
These points will be discussed in details in this Section.

The first step in the procedure consists in determining the region of convergence of the Fox function $\mrH$ and then 
we are left with the computation of $\mrH$. To start with we consider the univariate case~\cite{HTF},
\bq
\mrH\,\Bigl[ \mrz\,;\,
 \lpar {\mathbf a}\,,\,{\mathbf A} \rpar\,;\,
 \lpar {\mathbf b}\,,\,{\mathbf B} \rpar\,;\,
 \lpar {\mathbf c}\,,\,{\mathbf C} \rpar\,;\,
 \lpar {\mathbf d}\,,\,{\mathbf D} \rpar \Bigr] =
\int_{\mrL} \frac{\mrd \mrs}{2\,\mri\,\pi}
\frac{
 \prod_{\mrj=1}^{\mrm}\,\eG{\mra_\mrj + \mrA_\mrj\,\mrs}\,
 \prod_{\mrj=1}^{\mrn}\,\eG{\mrb_\mrj - \mrB_\mrj\,\mrs}
    } 
    {
 \prod_{\mrj=1}^{\mrp}\,\eG{\mrc_\mrj + \mrC_\mrj\,\mrs}\,
 \prod_{\mrj=1}^{\mrq}\,\eG{\mrd_\mrj - \mrD_\mrj\,\mrs}
     }\,\mrz^{\mrs} \spc
\label{UFFtwo}
\eq
where $\eG{\mrz}$ is the Euler Gamma function~\cite{HTF}.
It is assumed that all the $\mrA_\mrj\,\dots\,\mrD_\mrj$ are real and positive, and that the path of integration 
is a straight line parallel to the imaginary axis  (with indentations, if necessary) to avoid the poles of
the integrand. 

Using the definition of \eqn{UFFtwo} we introduce
\bqa
\ovA = \sum_{\mri=1}^{\mrm}\,\mrA_\mrj &\qquad \dots \qquad& \ovD= \sum_{\mrj=1}^{\mrq}\,\mrD_\mrj \spc
\nl  
\ova = \Re \sum_{\mri=1}^{\mrm}\,\mra_\mrj &\qquad \dots \qquad& \ovd= \Re \sum_{\mrj=1}^{\mrq}\,\mrd_\mrj \spc
\eqa
and define the following parameters:
\bqa
\alpha &=& \ovA + \ovB - \ovC - \ovD \spc \qquad
\beta  = \ovA - \ovB - \ovC + \ovD \spc 
\nl
\lambda &=& \frac{1}{2}\,\lpar \mrp + \mrq - \mrm - \mrn \rpar + \ova + \ovb - \ovc - \ovd \spc \qquad
\rho =  
\prod_{\mrj=1}^{\mrm}\,\mrA_{\mrj}^{\mrA_\mrj}\, 
\prod_{\mrj=1}^{\mrn}\,\mrB_{\mrj}^{ - \mrB_\mrj}\, 
\prod_{\mrj=1}^{\mrp}\,\mrC_{\mrj}^{ - \mrC_\mrj}\, 
\prod_{\mrj=1}^{\mrq}\,\mrD_{\mrj}^{\mrD_\mrj} \spp
\label{cpar}
\eqa 
There is a second definition of the Fox function given in \Bref{compH}:
\bq
\mrH\,\Bigl[ \mrz\,;\,(\mra_1,\mrA_1)\,\dots\,(\mra_\mrp,\mrA_\mrp)\,;\,
(\mrb_1,\mrB_1)\,\dots\,(\mrb_\mrq,\mrB_\mrq) \Bigr] =
\int_{\mrL} \frac{\mrd \mrs}{2\,\mri\,\pi}
\frac{
\prod_{\mrj=1}^{\mrm}\,\eG{\mrb_\mrj + \mrB_\mrj\,\mrs}\,\prod_{\mrj=1}^{\mrn}\,\eG{1 - \mra_\mrj - \mrA_\mrj\,\mrs}
     }
     {
\prod_{\mrj=\mrm+1}^{\mrq}\,\eG{1 - \mrb_\mrj - \mrB_\mrj\,\mrs}\,
\prod_{\mrj=\mrn+1}^{\mrp}\,\eG{\mra_\mrj + \mrA_\mrj\,\mrs}
     }\,\mrz^{ - \mrs} \spp
\label{UFFthree}
\eq
In \Bref{compH} the following parameters are defined:
\bqa
\mu &=& \sum_{\mrj=1}^{\mrq}\,\mrB_\mrj - \sum_{\mrj=1}^{\mrp}\,\mrA_\mrj \spc \qquad
\delta = \sum_{\mrj=1}^{\mrq}\,\mrb_\mrj - \sum_{\mrj=1}^{\mrp}\,\mra_\mrj + \frac{1}{2}\,(\mrp - \mrq) \spc
\nl
\beta &=& \Bigl[ \prod_{\mrj=1}^{\mrp}\,\mrA_\mrj^{ - \mrA_\mrj} \Bigr]\,
          \Bigl[ \prod_{\mrj=1}^{\mrq}\,\mrB_\mrj^{\mrB_\mrj} \Bigr] \spc \qquad
\alpha = \sum_{\mrj=1}^{\mrn}\,\mrA_\mrj -
         \sum_{\mrj=\mrn+1}^{\mrp}\,\mrA_\mrj +
         \sum_{\mrj=1}^{\mrm}\,\mrB_\mrj -
         \sum_{\mrj=\mrm+1}^{\mrq}\,\mrB_\mrj \spp
\eqa

It is important to observe that in \Bref{compH} the integration contour separates the poles of 
$\eG{\mrb_\mrj + \mrB_\mrj\,\mrs}$ from the poles of $\eG{1 - \mra_\mrj - \mrA_\mrj\,\mrs}$.

There are three different paths of integration; according to \Bref{HTF} we have:
\bei

\item[\ovalbox{$\mrL_{\mri\,\infty}$}] $\mrL$ runs from $-\,\mri\,\infty$ to $+\,\mri\,\infty$ separating the poles
of the integrand.

\item[\ovalbox{$\mrL_{+\,\infty}$}] $\mrL$ is a loop starting and ending at $+\,\infty$ and encircling all poles of
$\eG{\mrb_\mrj - \mrs}$ once in the negative direction.

\item[\ovalbox{$\mrL_{-\,\infty}$}] $\mrL$ is a loop starting and ending at $-\,\infty$ and encircling all poles of
$\eG{1 - \mra_\mrj + \mrs}$ once in the positive direction.

\eei
It is assumed that at least one of the three definitions makes sense. In cases where more than one make sense, they
lead to the same result.

Using the notations of \eqn{UFFtwo} with $\phi = \marg(\mrz)$ and $\mrs = \sigma + i\,\mrt$ we have the four different 
cases described in \Bref{HTF} and repeated here for the benefit of the reader:
\bei

\item[\ovalbox{I}] $\alpha > 0$. The integral converges absolutely for $\mid \phi \mid < \alpha \pi/2$ (the
point z = 0 is tacitly excluded.).

\item[\ovalbox{II}] 
$\alpha = 0, \beta \not= 0$. The integral does not converge for
complex $\mrz$. For $\mrz > 0$ it converges absolutely if $\sigma$ is so chosen that
$\beta\,\sigma + \lambda < - 1$
and there exists an analytic function of $\mrz$, defined over $\mid \phi \mid < \pi$ 
whose values for positive $\mrz$ are given by $\mrH$ defined in \eqn{UFFtwo}.

\item[\ovalbox{III}] $\alpha = \beta = 0$, $\lambda < -1$
The integral converges absolutely for all positive $\mrz$ (but not for complex $\mrz$)
and represents a continuous function of $\mrz$ ($0 < \mrz < \infty$ ). There are now two
analytic functions, one regular in any domain contained in $\mid \phi \mid < \pi$, $\mid \mrz \mid > \rho$
whose values for $\mrz > \rho$ are represented by $\mrH$, and another regular
in any domain contained in $\mid \phi \mid < \pi$, $0 < \mid \mrz \mid < \rho$ whose values for
$0 < \mrz < \rho$ are represented by $\mrH$. The two functions are in general distinct.

\item[\ovalbox{IV}]
$\alpha = \beta = 0$, $- 1 < \lambda < 0$.
The integral converges (although not absolutely) for $0 < \mrz < \rho$ and for $\mrz > \rho$. There are two analytic 
functions of the same nature as in the preceding case. There is a discontinuity
at $\mrz = \rho$ and the integral does not exist there, though it may have a principal
value. 

\eei
A special case of interest is $\alpha > 0$ and $\phi = \alpha \pi/2$. Introducing $\mrs = \sigma + i\,\mrt$ the 
large $\mid \mrt \mid$ behavior of the integrand is controlled by
\bq
\mid \mrt \mid^{\beta\,\sigma + \lambda} \spc
\eq
requiring $\beta\,\sigma + \lambda < - 1$. According to \Bref{compH} we have the following cases:
\begin{enumerate}

\item If $\mu > 0$ or $\mu = 0$ and $\mid \mrz \mid < \beta$ then $\mrL = \mrL_{-\,\infty}$;

\item If $\mu < 0$ or $\mu = 0$ and $\mid \mrz \mid > \beta$ then $\mrL = \mrL_{+\,\infty}$;

\item If $\alpha > 0$ and $\marg(\mrz) < 1/2\,\alpha\,\pi$ or $\alpha = 0$ and
$\beta\,\sigma + \Re\,\delta < - 1$ then we can use $\mrL = \mrL_{\mri\,\infty}$.

\end{enumerate}
The most general, multivariate Fox function has been introduced in \Bref{HS} (see also \Brefs{HY,BS}),
which gives the most general criterion to determine the regions of convergence of a multivariate Fox function;
given ${\mathbf s} = [\mrs_1\,\dots\,\mrs_{\mrr} ]$,
${\mathbf \alpha} = [\alpha_1\,\dots\,\alpha_{\mrr} ]$,
${\mathbf \beta} = [\beta_1\,\dots\,\beta_{\mrr} ]$,
$\marg({\mathbf z}) = [\marg(\mrz_1)\,\dots\,\marg(\mrz_{\mrr}) ]$, we define
\bq
{\mathbf A} = \lpar \mra_{\mrj , \mrk} \rpar_{\mrm \times \mrr} \spc \qquad
{\mathbf B} = \lpar \mrb_{\mrj , \mrk} \rpar_{\mrn \times \mrr} \spc
\eq
\bq
\mrH\Bigl[ {\mathbf z}\,;\,({\bf \alpha}\,,\,{\mathbf A} )\,;\,({\bf \beta}\,,\,{\mathbf B}) \Bigr] =
\Bigl[ \prod_{\mrj=1}^{\mrr}\,\int_{\mrL_\mrj}\,\frac{\mrd \mrs_\mrj}{2\,\mri\,\pi} \Bigr]\,
\Uppsi\,\,\prod_{\mrj=1}^{\mrr}\,( \mrz_\mrj)^{- \mrs_\mrj} \spc \quad
\Uppsi = \frac{ \prod_{\mrj=1}^{\mrm}\,\eG{\alpha_{\mrj} + \sum_{\mrk}\,\mra_{\mrj , \mrk}\,\mrs_{\mrk}}}
            { \prod_{\mrj=1}^{\mrn}\,\eG{\beta_{\mrj} + \sum_{\mrk}\,\mrb_{\mrj , \mrk}\,\mrs_{\mrk}}} \spc
\label{gmff}
\eq
where $\mra$ and $\mrb$ are arbitrary real numbers.
It is important to realize that the multiple integral in \eqn{gmff} may be overall divergent although the iterate 
integrals converge~\cite{HS}. The convergenge problem has been discussed in \Bref{HS}.
Once the MB representation of a Feynman integral has been obtained, the object to be computed is
\bq
\mrH = \Bigl[ \prod_{\mrj=1}^{\mrr}\,\int_{\mrL_\mrj}\,\frac{\mrd \mrs_\mrj}{2\,\mri\,\pi} \Bigr]\,
\mrF\lpar \mrs_1\,\dots\,\mrs_\mrr \rpar\,\prod_{\mrj=1}^{\mrr}\,\mrz_\mrj^{\mrs_\mrj} \spp
\label{MBinta}
\eq
The function $\mrF$ is a quotient of products of Euler Gamma functions;
our strategy is that the paths of integration are straight lines parallel to the imaginary axis 
avoiding the poles of the integrand~\cite{HTF}. 
Therefore, introducing $\mrs_\mrj = \sigma_\mrj + i\,\mrt_\mrj$, we can rewrite \eqn{MBinta} as
\bq
\mrH = \Bigl[ \prod_{\mrj=1}^{\mrr}\,\int_{- \infty}^{+ \infty}\,\frac{\mrd \mrt_\mrj}{2\pi} \Bigr]\,
\mrF\lpar \mrs_1\,\dots\,\mrs_\mrr \rpar\,\prod_{\mrj=1}^{\mrr}\,\mrz_\mrj^{\mrs_\mrj} \spc
\label{MBintb}
\eq
and we proceed with a numerical computation of $\mrH$.
There are two strategies in deriving the MB representation; we can use a MB splitting,
\bq
\Bigl[ \prod_{\mrj=0}^{\mrn} \mrA_\mrj \Bigr]^{ - \alpha} =
\iMB{\mrn}\,\Bigl[ \prod_{\mrj=1}^{\mrn}\,\eG{ - \mrs_\mrj} \Bigr]\,
\frac{\eG{\alpha + \sum_{\mrj}\,\mrs_\mrj}}{\eG{\alpha}}\,
\mrA_0^{-\alpha - \sum\,\mrs_\mrj}\,\prod_{\mrj=1}^{\mrn}\,\mrA_\mrj^{\mrs_\mrj} \spp
\eq
Alternatively we can use the Feynman parametrization and perform the first integral obtaining a
(generalized) hypergeometric function (usually an $\mrF^{(\mrN}_{\mrD}$ Lauricella function~\cite{FDMB}). Using the
MB representation of the result allows us to compute the second integral and we can repeat the procedure
until we obtain the final result.
Let us consider what happens when computing Feynman integrals in the physical region. The integrals
have a well{-}defined cut structure fixed by the solution of the corresponding Landau equations once the
Feynman prescription has been taken into account. If the MB contour(s) is $\mrL_{\pm\,\infty}$ we compute
the residues of the poles; however, unless we are able to sum the residues, it will not be easy to see
the cut structure. If the contour(s) is $\mrL_{\mri\,\infty}$ we can do a numerical integration and try
to understand the correctness of the resulting imaginary part of the result.
In other words, our goal is to define Feynman integrals below and above thresholds as Mellin{-}Barnes
integrals with contour(s) $\mrL_{\mri\,\infty}$ in order to proceed with a numerical integration~\cite{NCFF}.
\paragraph{Mellin{-}Barnes splitting} \hspace{0pt} \\
Consider a simple example of the MB splitting:
\bq
\mrF = ( \mrQ + \mra - \mri\,\delta)^{-\rho} \spc
\eq
where $\mrQ$ is a function of Feynman parameters and $\mra$ is a positive
parameter, with $\delta \to 0_{+}$. We perform a MB splitting, i.e.
\bq
 \mrF = \mra^{-\rho}\,\intMB\,\eB{\mrs}{\rho - \mrs}\,\Bigl(
\frac{\mra}{\mrQ - \mri\,\delta} \Bigr)^{\mrs} \spc
\eq
where $\mrB$ is the Euler Beta function and $0 < \Re\,s < \rho$. The
choice of $\mrL$ depends on
$\mrz = \mra/\mrQ$. Indeed $\mrF$ is proportional to a Meijer
$\mrG^{1,1}_{1,1}$
function with parameters $\mra_1 = 1$ and $\mrb_1 = \rho$.

If $\mid \mrz \mid < 1$ we select $\mrL = \mrL_{+\,\infty}$; if $\mid \mrz
\mid > 1$ we select $\mrL = \mrL_{-\,\infty}$ and compute the residues of
the poles.

The Meijer $\mrG$ function~\cite{HTF} has parameters $\alpha = 2, \beta = 0$ and $\lambda = \rho > 0$, so that 
the MB integral over $\mrL_{\mri\,\infty}$ does not converge if $\mrQ < 0$, despite the $ -\,\mri\,\delta$ prescription.
The main question will be how to use $\mrL = \mrL_{\mri\,\infty}$. Few simple examples will follow, 
illustrating the procedure. Given
\bq
\mrI = \int_0^1 \frac{\mrd \mrx}{\mra - \mrx} = - \ln\Bigl( 1 - \frac{1}{\mra} \Bigr) 
\spc\qquad \mra > 0, \spc
\quad \mra \to \mra - \mri\delta \spp
\eq
we write
\bq
 \mrI = \frac{1}{\mra}\,\intMB\,\Gamma(\mrs)\,\Gamma(1 - \mrs)\,\Bigl( -
\frac{\mrx}{\mra}\Bigr)^{-\mrs} \spc
 \eq
we have the following scenarios:
\begin{itemize}
    \item[]
    $\mra > 1$ \spc \qquad $\mid - \frac{\mrx}{\mra} \mid < 1$ \spc

    \item[] $0 < \mra < 1$ \spc \quad $\mrx < \mra$ \spc \quad $\mid -
\frac{\mrx}{\mra} \mid < 1$

    \item[]
     $0 < \mra < 1$ \spc \quad $\mrx > \mra$ \spc \quad $\mid - \frac{\mrx}{\mra} \mid  > 1$
\end{itemize}
Therefore, with $\Re\,\mrs < 1$, we obtain
\bq
\int_0^\mra \mrd \mrx\,
\mrx^{-\mrs} = \frac{\mra^{1 - \mrs}}{1 - \mrs} 
\spc \qquad
\int_\mra^1 \mrd \mrx\,\mrx^{-\mrs} = \frac{1}{1 - \mrs}\,
(1 - \mra^{1 - \mrs}) \spp
\eq
Therefore the corresponding MB representation will be $\mrI = \mrI_{-} +
\mrI_{+}$ where
\bq
 \mrI_{-} = \intMB\,\frac{\Gamma(\mrs)\,\Gamma(1 - \mrs)}{1 - \mrs}\,( -
\mra)^{\mrs}\,\mra^{1 - \mrs} \spc \quad
  \mrI_{+} = \intMB\,\frac{\Gamma(\mrs)\,\Gamma(1 - \mrs)}{1 - \mrs} \,( -
\mra)^{\mrs}\,( 1 - \mra^{1 - \mrs}) \spp
\eq
If we do the $\mrx$ integral without selecting the $\mrL$ contour we obtain
\bq
  \mrI = \frac{1}{\mra}\,\intMB\,
  \frac{\Gamma(s)\,\Gamma^2(1 - s)}{\Gamma(2 - s)}\,( - \mra)^{\mrs} \spc
\eq
and we obtain a Mejier $\mrG$ function with parameter~\cite{compH} $\mrm =
1, \mrn = 2$ and $\mrp = \mrq = 2$ giving $\mu = 0$ and $\beta = 1$.
Therefore, for $\mra > 1$ we must select $\mrL = \mrL_{-\,\infty}$ obtaining
\bq
\mrI = \sum_{\mrm=0}^{\infty}\,\frac{\mra^{-1 -m}}{\mrm +1} = - \ln\Bigl(1
- \frac{1}{\mra}\Bigr) \spp
\eq
When $0 < \mra < 1$ we must select $\mrL = \mrL_{+\,\infty}$ where the
relevant poles are at $\mrs = 1$ (double pole) and at $\mrs= \mrm + 1,
\mrm
\ge 1$ (single poles). The result is
\bq
   \hbox{single poles} \quad \mapsto \quad - \ln(1 - \mra) \spc \qquad \hbox{double
pole} \quad \mapsto \quad - \ln \bigl( - \frac{1}{\mra} \bigr) \spp
\eq
The conclusion is that we can perform the $\mrx$ integration before
selecting $\mrL$, as long as $\mrL_{\pm\,\infty}$ intersect the real axis
at $0 < \Re\,\mrs < 1$.

An important question is related to the correct definition of
$\mrL_{\pm\,\infty}$. Consider the following example:
\bq
   \mrI = \int_0^1 \mrd \mrx\,(1 + \mra\,\mrx)^{\alpha} = \frac{1}{1 +
\alpha}\,\frac{1}{\mra}\,\Bigl[ (1 + \mra)^{\alpha + 1} - 1 \Bigr] \spp
\eq
We can write
\bq
  \mrI = \frac{1}{\Gamma( - \alpha)}\,\intMB\,\Gamma( - \mrs)\,\frac{\Gamma( - \alpha + \mrs)}{1 + \mrs} \,\mra^{\mrs}
\eq
If $\mra < 1$ we select $\mrL = \mrL_{+\,\infty}$
and we have poles at $\mrs = \mrm_1$, $\mrs = - 1$
(double pole) and additional poles at $\mrs = - 1 - \mrm_2, \mrm_2 > 1$
and $\mrs= \alpha - \mrm_3$, The correct result follows if
$\mrL_{+\,\infty}$ crosses the real axis at
$ - 1 < \Re\,\mrs < 0$ but indented to avoid the pole at $\mrs = \alpha$.

In the following we give additional examples.
\paragraph{Example $1$:  hypergeometric functions} \hspace{0pt} \\
Consider now the following Euler{-}Mellin integral:
\bq
\mrI = \int_0^1 \mrd \mrx\,\mrx^{\mrb - 1}\,(1 - \mrx)^{\mrc - \mrb - 1}\,(1 - \mrz\,\mrx)^{ - \mra} \spp
\eq
If $\Re \mrc > \Re \mrb > 0$ and $\mid \marg(1 - \mrz) \mid < \pi$ we can write
\bq
\mrI = \eB{\mrb}{\mrc - \mrb}\,\hyp{\mra}{\mrb}{\mrc}{\mrz} \spc
\eq
where $\mrB$ is the Euler Beta function.
Note that with $\mrz \to \mrz - \mri\,\delta$ the original integral can be interpreted as a Hadamard finite{-}part 
integral~\cite{FPI,Hreg} even if $\mrz \in \Rf$ and $ \mrz > 1$. Next we would like to write $\mrI$ as a 
Mellin{-}Barnes integral, \ie
\bq
\mrI = \frac{\eG{\mrc - \mrb}}{\eG{\mra}}\,\int_{\mrL}\,\frac{\mrd \mrs}{\tip}
\,\eG{ - \mrs}\,\frac{\eG{\mra + \mrs}\,\eG{\mrb + \mrs}}{\eG{\mrc + \mrs}}\,( - \mrz )^\mrs \spc
\label{IMB}
\eq
which requires $\mid \marg( - \mrz) \mid < \pi$. Therefore, in order to write $\mrI$ in terms of a MB
integral and $\mrz \in \Rf$ we should start with $\mrz < 0$. If we consider the $\mrI$ function as
represented in \eqn{IMB} the corresponding parameters (with the conventions of \Bref{HTF})
are $\mrm = 2, \mrn = 1$ and $\mrp = 1, \mrq = 0$ with
$\alpha = 2, \beta = 0$ and
$\lambda = - 1 + \Re (\mra + \mrb - \mrc)$.
Therefore, as long as $\Re (\mra + \mrb - \mrc) < 0$ the MB integral converges for $\mrL = \mrL_{\mri\,\infty}$
even if $\mrz \in \Rf$ and $\mrz > 0$. However, it is easily seen that the analytic continuation does not
reproduce the cut structure of the original integral, namely we have a cut at $[ 0\,,\,\infty ]$ instead of
a cut at $[ 1\,,\,\infty ]$. Seen in terms of a Feynman integral this corresponds to the fact, with this
procedure, we can describe the integral above its normal threshold but not below it.

In order to understand the correct procedure we consider the following example:
\bq 
\mrI = \int_0^1 \mrd \mrx\,\mrx^{-1/2}\,(1 - \mrz\mrx)^{-1} = 2\;\hyp{1}{\frac{1}{2}}{\frac{3}{2}}{\mrz} \spc
\eq
with $\mrz \in \Rf$. There are three cases:
\bei

\item[\ovalbox{$\mrz < 0$}] Here we immediately obtain
\bq
\mrI = \int_{\mrL_{\mri\,\infty}}\,\frac{\mrd \mrs}{\tip}\,\eG{ - \mrs}\,
\frac{\eG{1 + \mrs}\,\eG{1/2 + \mrs}}{\eG{3/2 + \mrs}}\,( - \mrz )^\mrs \spp
\eq

\item[\ovalbox{$0 < \mrz < 1$}] We use the following transformation of the Gauss hypergeometric function,
\bq
\hyp{\mra}{\mrb}{\mrc}{\mrz} = (1 - \mrz)^{ - \mra}\,
\hyp{\mra}{\mrc - \mrb}{\mrc}{\frac{\mrz}{\mrz - 1}} \spc
\eq
where now the argument of the hypergeometric function is negative; we obtain
\bq
\mrI = \sqrt{\pi}\,(1 - \mrz)^{ - 1}\,
\int_{\mrL_{\mri\,\infty}}\,\frac{\mrd \mrs}{\tip}\,\eG{ - \mrs}\,
\frac{\eGs{1 + \mrs}}{\eG{3/2 + \mrs}}\,\lpar \frac{\mrz}{1 - \mrz} \rpar^{\mrs} \spp
\eq

\item[\ovalbox{$\mrz > 1$}] Here we use~\cite{HTF}
\bqa
\hyp{\mra}{\mrb}{\mrc}{\mrz} &=& - \Bigl[ \mrc\,(\mrc + 1)\,(\mrz - 1) \Bigr]^{-1}\,
 \Bigl\{
  (\mrc + 1)\,\Bigl[ \mrc - (2\,\mrc - \mra - \mrb + 1)\,\mrz \Bigr]\,
\hyp{\mra}{\mrb}{\mrc + 1}{\mrz}
\nl
{}&+&
(\mrc - \mra + 1)\,(\mrc - \mrb + 1)\,\mrz\;
\hyp{\mra}{\mrb}{\mrc + 2}{\mrz} \Bigr\} \spc
\eqa
and obtain the following MB representation:
\bqa
\mrI &=& \frac{5\,\mrz - 3}{2\,(\mrz - 1)}\,\int_{\mrL_{\mri\,\infty}}\,\frac{\mrd \mrs}{\tip}\,
\eG{ - \mrs}\frac{\eG{1 + \mrs}\,\eG{1/2 + \mrs}}{\eG{5/2 + \mrs}}\,( - \mrz )^{\mrs}
\nl
{}&-& 3\,\frac{\mrz}{\mrz - 1}\,\int_{\mrL_{\mri\,\infty}}\,\frac{\mrd \mrs}{\tip}\,
\eG{ - \mrs}\frac{\eG{1 + \mrs}\,\eG{1/2 + \mrs}}{\eG{7/2 + \mrs}}\,( - \mrz )^{\mrs} \spc
\eqa
where now the two MB integrals are convergent even for $\mrz > 1$.

\eei
\paragraph{Example $2$: Lauricella functions} \hspace{0pt} \\
Another important result follows from considering
\bq
\mrI = \int_0^1\,\mrd \mrx \lpar \lambda - \mrx + \mrx^2 \rpar^{ - \alpha} \spc
\label{fexa}
\eq
which for $\lambda = \mrm^2/\mrs$ is proportional to a two{-}point function in arbitrary space{-}time
dimensions. The normal threshold is at $\lambda = 1/4$; clearly we could write
\bq
\mrI = \lambda^{ - \alpha}\,
\LFD{1}{\alpha}{\alpha}{2}{\frac{1}{\mrx_{-}}}{\frac{1}{\mrx_{+}}} \spc
\eq
where $\mrF^{(2)}_{\mrD}$ is a Lauricella function~\cite{FDMB} and
where we have introduced the two roots of the quadratic form appearing in \eqn{fexa}. However, the best
strategy is to write the integral in terms of a parameter $\beta$, with $\beta^2 = 1 - 4\,\lambda$,
\bqa
\mrI &=& \int_0^1\,\mrd \mrx\,\Bigl[ ( \mrx - \frac{1}{2} )^2 - \frac{1}{4}\,\beta^2 \Bigr]^{ - \alpha} = 
\eB{\frac{1}{2}}{\alpha - \frac{1}{2}}\,\lpar - \frac{1}{4}\,\beta^2 \rpar^{1/2 - \alpha} 
\nl
{}&-&
\,\frac{2^{3 + 2\,\alpha}}{\alpha - \frac{1}{2}}\;
\hyp{\alpha}{\alpha - \frac{1}{2}}{\alpha + \frac{1}{2}}{\beta^2} \spp
\eqa
If $ 0 < s < 4\,\mrm^2$ it follows $\beta^2 < 0$ and we can safely use the MB representation of the
hypergeometric function. Indeed the imaginary part arises according to
\bq
\mrH = \int_{ - \infty}^{+ \infty}\,\mrd \mrt\,\mrR(\sigma + \mri\,\mrt)\,\mrz^{\sigma + \mri\,\mrt} \spc
\eq
where $\mrR$ is a quotient of products of Gamma functions and $\ln \mrz = \mrL + \mri\,\upphi$. Then
\bq
\mrH = 2\,\exp\{\sigma\,(\mrL + \mri\,\upphi)\}\,\int_0^{\infty}\,\mrd \mrt\,
      \Bigl[ \cosh(\mrt\,\uppi)\,\Re \lpar \mrR\,e^{\mri\,\mrL\,\mrt} \rpar  +
\mri\,\sinh(\mrt\,\uppi)\,\Im \lpar \mrR\,e^{\mri\,\mrL\,\mrt} \rpar \Bigr] \spc
\eq
where we have used $\mrR(\mrz^{*})= \mrR^{*}(\mrz)$.

A more general case is as follows: imagine that at a certain stage of the calculation of a Feynman integral we
end up with a Lauricella function
\bq
\mrI = \LFD{\mra}{\mrb_1}{\mrb_2}{\mrc}{\mrx}{\mry} \spc
\eq
and we want to write the MB representation when $\mrx, \mry \in \Rf$. We have to consider the following 
cases~\cite{HTF,Asur}:
\begin{enumerate}

\item \ovalbox{$\mrx < 0, \mry < 0$}. We can use
\bq
\mrF^{(2)}_{\mrD} = \iMB{2}\,\frac{\eG{\mra + \mrs_1 + \mrs_2}}{\eG{\mrc + \mrs_1 + \mrs_2}}\,
\Bigl[ \prod_{\mrj=1}^{2}\,\eG{ - \mrs_\mrj}\,\eG{\mrb_\mrj + \mrs_\mrj} \Bigr]\,
( - \mrx)^\mrs_1\,( - \mry)^{\mrs_2} \spp
\eq

\item \ovalbox{$0 < \mrx < 1, \mry < 0$}. First we use the transformation~\cite{HTF,Asur}
\bq
\mrI = (1 - \mrx)^{ - \mra}\,
\LFD{\mra}{\mrc - \mrb_1 - \mrb_2}{\mrb_2}{\mrc}{\frac{\mrx}{\mrx - 1}}{\frac{\mry - \mrx}{1 - \mrx}} \spc
\eq
so that both arguments are negative and we can use the standard MB representation.

\item \ovalbox{$0 < \mrx < 1, 0 < \mry < 1$}. First we use the transformation~\cite{HTF,Asur}
\bq
\mrI = (1 - \mrx)^{ - \mrb_1}\,(1 - \mry)^{ - \mrb_2}\,
\LFD{\mrc - \mra}{\mrb_1}{\mrb_2}{\mrc}{\frac{\mrx}{\mrx - 1}}{\frac{\mry}{\mry - 1}} \spp
\eq

\item \ovalbox{$0 < \mrx < 1, \mry > 1$}. We transform variables and obtain
\bq
\mrI = ( 1 - \mrx)^{ - \mra}\,\iMB{2}\,\frac{\eG{\mra + \mrs_1 + \mrs_2}}{\eG{\mrc + \mrs_1 + \mrs_2}}\,
\Bigl[ \prod_{\mrj=1}^{2}\,\eG{ - \mrs_\mrj}\,\eG{\mrb_\mrj + \mrs_\mrj} \Bigr]\,
\lpar \frac{\mrx}{1 - \mrx} \rpar^{\mrs_1}\,\lpar \frac{\mrx - \mry}{1 - \mrx} \rpar^{\mrs_2} \spp
\eq
The $\mrs_2$ integral corresponds to parameters $\alpha = 2, \beta = 0$ and
$\lambda = - 1 + \Re(\mra + \mrb_2 - \mrc)$. If $\lambda < - 1$ we can use the standard MB representation; if
not we continue by using the following contiguity relation
\bq
\mrc\,\LFD{\mra}{\mrb_1}{\mrb_2 - 1}{\mrc}{\mrx}{\mry} +
\mrc\,(\mry - 1)\,\LFD{\mra}{\mrb_1}{\mrb_2}{\mrc}{\mrx}{\mry} -
(\mrc - \mra)\,\mry\,\LFD{\mra}{\mrb_1}{\mrb_2}{\mrc + 1}{\mrx}{\mry} = 0 \spc
\label{conty}
\eq 
until we obtain $\lambda < - 1$, so that the standard MB representation can be used.

\item \ovalbox{$\mrx < 0, \mry > 1$}. The procedure is the same as in the previous case.

\item \ovalbox{$\mrx > 1, \mry < 0$}. In this case $\lambda = - 1 + \Re(\mra + \mrb_1 - \mrc)$ for the $\mrs_1$ integral. 
If needed we use~\cite{Asur}
\bq
\mrc\,\LFD{\mra}{\mrb_1 - 1}{\mrb_2}{\mrc}{\mrx}{\mry} +
\mrc\,(\mrx - 1)\,\LFD{\mra}{\mrb_1}{\mrb_2}{\mrc}{\mrx}{\mry} -
(\mrc - \mra)\,\mrx\,\LFD{\mra}{\mrb_1}{\mrb_2}{\mrc + 1}{\mrx}{\mry} = 0 \spc
\label{contx}
\eq 

\item \ovalbox{$\mrx >1, \mry > 1$}. If needed both \eqn{contx} and \eqn{conty} can be used

\end{enumerate}

The strategy discussed so far can be generalized to deal with arbitrary $\mrF^{(\mrN)}_{\mrD}$ functions; suppose
we have to deal with
\bq
\mrF^{(\mrN)}_{\mrD}\lpar \mra\,;\,\mathbf{b}\,;\,\mrc\,;\,\mathbf{z} \rpar \spc
\eq
where $\mathbf{b} = \mrb_1\,,\,\dots\,,\,\mrb_{\mrN}$, and $\mrz_1 \in \Rf$ with $0 < \mrz_1 < 1$. 
Let us define $\mathbf{z}^{\prime} = \mrz_2\,,\,\dots\,,\,\mrz_{\mrN}$ \etc
Under the condition $\mid \mrz_\mrj \mid < 1,\;\mrj=2,\dots,\mrN$ we can write
\bq
\mrg\lpar \mra\,;\,\mathbf{b}\,;\,\mathbf{z}\,;\,\mrc\,;\,\mrs \rpar =
\mrg_0\,(1 - \mrz_1)^{\mrs}\,
\mrF^{(\mrN-1)}_{\mrD}\lpar \mra + \mrs\,;\,\mathbf{b}^{\prime}\,;\,\mrc - \mrb_1\,;\,\mathbf{z}^{\prime} \rpar \spc
\eq
\bq
\mrg_0 = \eG{ - \mrs}\,\eG{\mra + \mrs}\,\eG{\mrb_1 + \mrs}\,\eG{\mrc - \mra - \mrb_1 - \mrs} \spc
\eq
and derive~\cite{FDMB} 
\bq
\mrF^{(\mrN)}_{\mrD}\lpar \mra\,;\,\mathbf{b}\,;\,\mrc\,;\,\mathbf{z} \rpar =
\frac{\eG{\mrc}}{\eG{\mra}\,\eG{\mrb_1}\,\eG{\mrc - \mra}\,\eG{\mrc - \mrb_1}}\,
\int_{\mrL}\,\frac{\mrd \mrs}{\tip}\,\mrg \spp
\eq
The transformations in the general case are given by:
\bqa
{}&{}&\mrF^{(\mrN)}_{\mrD}(\mra\,;\,\mathbf{b}\,;\,\mrc\,;\,\mrx\,,\,\mry_1\,\dots\,\mry_{\mrN-1}) =
(1 - \mrx)^{ - \mra}
\nl
{}&\times&
\mrF^{(\mrN)}_{\mrD}\Bigl( \mra\,;\,\mrc - \sum_{\mrj}\,\mrb_\mrj\,,\,\mrb_2\,\dots\,\mrb_{\mrN}\,;\,\mrc\,;\,
\frac{\mrx}{\mrx - 1}\,,\,\frac{\mry_1 - \mrx}{1 - \mrx}\,\dots\,\frac{\mry_{\mrN-1} - \mrx}{1 - \mrx} \Bigr) \spc
\eqa
\bqa
{}&{}&\mrF^{(\mrN)}_{\mrD}(\mra\,;\,\mathbf{b}\,;\,\mrc\,;\,\mrx\,,\,\mry_1\,\dots\,\mry_{\mrN-1}) =
(1 - \mrx)^{\mrc - \mra - \mrb_1}\,\prod_{\mrj=2}^{\mrN}\,(1 - \mry_{\mrj - 1})^{ - \mrb_\mrj}
\nl
{}&\times&
\mrF^{(\mrN)}_{\mrd}\Bigl(\mrc - \mra\,;\,\mrc - \sum_{\mrj}\,\mrb_\mrj\,,\,\mrb_2\,\dots\,\mrb_{\mrN}\,;\,\mrc\,;\,
\mrx\,,\,\frac{\mrx - \mry_1}{1 - \mry_1}\,\dots\,\frac{\mrx - \mry_{\mrN-1}}{1 - \mry_{\mrN-1}} \Bigr) \spp
\eqa
The first transformation can be used when we have a Lauricella function $\mrF^{(\mrN)}_{\mrD}$ of variables
$\mrz_1\,\dots\,\mrz_{\mrN}$ with
\bq
0 < \mrz_\mrj < 1 \spc \quad \mrj=1\,\dots\,\mrM \spc \qquad
\mrz_\mrj > 1 \spc \quad \mrj= \mrM + 1\,\dots\,\mrN \spc
\eq
Assuming that $\mrz_1 = \max_{\mrj=1\,\dots\,\mrM}\,\{\mrz_\mrj\}$ the transformation gives
\bq
\frac{\mrz_1}{\mrz_1 - 1}\,,\,\frac{\mrz_2 - \mrz_1}{1 - \mrz_1}\,\dots\,\frac{\mrz_{\mrM} - \mrz_1}{1 - \mrz_1} \spc
\eq
where all variables are negative and
\bq
\frac{\mrz_{\mrM + 1} - \mrz_1}{1 - \mrz_1}\,\dots\,\frac{\mrz_{\mrN} - \mrz_1}{1 - \mrz_1} \spc
\eq
where all variables are greater than one.
\paragraph{Example $3$} \hspace{0pt} \\
Consider the following example:
\bq
\mrH = \int_0^1\,\Bigl[ \prod_{\mrj=1}^{2}\,\mrd \mrx_\mrj\,\mrx_{\mrj}^{\mra_\mrj-1}\,
(1 - \mrx_\mrj)^{\mrb_\mrj - \mra_\mrj - 1} \Bigr]\,(1 - \mrz\,\mrx_1\,\mrx_2)^{- \mra_3} \spp
\eq
To compute the integral we can perform a MB splitting:
\bq
(1 - \mrz\,\mrx_1\,\mrx_2)^{- \mra_3} = \int_{\mrL}\,\frac{\mrd \mrs}{2\,\mri\,\pi}\,
\eB{\mrs}{\mra_3 - \mrs}\,\Bigl( - \frac{1}{\mrz\,\mrx_1\,\mrx_2} \Bigr)^{\mrs} \spc
\eq
and use
\bq
\int_0^1\,\mrd \mrx_\mrj\,\mrx_{\mrj}^{\mra_\mrj-1-\mrs}\,
(1 - \mrx_\mrj)^{\mrb_\mrj - \mra_\mrj - 1} = \eB{\mra_\mrj - \mrs}{\mrb_\mrj - \mra_\mrj} \spc
\eq
which requires $\mra_\mrj  > \Re\,\mrs$ for $\mrj= 1,2$. Assuming $\mrz \in \Rf$ and $\mid \mrz \mid > 1$ we
derive
\bq
   \mrI =
   \frac{
   \prod_{\mrj=1}^{2}\,\Gamma(\mrb_\mrj -
\mra_\mrj)}{\Gamma(\mra_3)}\,\int_{\mrL}\,\frac{\mrd
\mrs}{2\,\mri\,\pi}\,\Gamma(\mrs)\,
   \Gamma(\mra_3 - \mrs)\,\prod_{\mrj=1}^{2}\,
   \frac{\Gamma(\mra_\mrj - \mrs)}{\Gamma(\mrb_\mrj - \mrs)}\,( - \mrz)^{- \mrs} \spp
\eq
According to the conventions of \Bref{compH} the MB integral corresponds to
parameters $\mrm = 1, \mrn = 3$ and $\mrp = \mrq = 3$, giving $\mu = 0$
and $\beta = 1$. Therefore for $\mid \mrz \mid > 1$ we have $\mrL =
\mrL_{-\,\infty}$ while for $\mid \mrz \mid < 1$ we have $\mrL =
\mrL_{+\,\infty}$.
The poles are at $\mrs = -\mrm_0$ (B poles) and at $\mrs = \mra_\mrj +
\mrm_\mrj$ (A poles). Alternatively we can write
\bq
 \mrI = \Bigl[\prod_{\mrj=1}^{2}\, \mathrm{B}(\mra_\mrJ\,,\,\mrb_\mrj -
\mra_\mrj) \Bigr]\,
 \hyptt(\mathbf{a}\,;\,\mathbf{b}\,;\,\mrz) \spc
\eq
requiring $\mid \mathrm{arg}(1 - \mrz) ) \mid < \pi$. If $\mrz \in \Rf$,
$\mrz < 1$ is required. We can also write
\bq
\hyptt(\mathbf{a}\,;\,\mathbf{b}\,;\,\mrz) =
\frac{\prod_{\mrj=1}^{2}\,\Gamma(\mrb_\mrj)}
{\prod_{\mrj=1}^{3}\,\Gamma(\mra_\mrj)}\,
\int_{\mrL_{\mri\,\infty}}\,\frac{\mrd \mrs}{2\,\mri\,\pi}\,\Gamma( - \mrs)\,
\frac{\prod_{\mrj=1}^{3}\,\Gamma(\mra_\mrj +
\mrs)}{\prod_{\mrj=1}^{2}\,\Gamma(\mrb_\mrj + \mrs)}\,( - \mrz )^{\mrs} \spc
\eq
which requires $\mid \mathrm{arg}( - \mrz) \mid < \pi$ or $\mrz < 0$ for
$\mrz \in \Rf$.
Once again the problem in writing a MB representation respecting the
correct imaginary part of the original integral is represented by
(generalized) hypergeometric functions of argument $0 < \mrz < 1$.

The first step will be to discuss the convergence of
the $\mrL_{\mri\,\infty}$ MB integral. According to the conventions of
\Bref{HTF} we have parameters $\mrm = 2, \mrn = 3$ and $\mrp = 0, \mrq = 2$
giving $\alpha = 2$ and $\lambda = - 1 - \uppsi_0$, where $\uppsi_0 =
\sum_\mrj\,\mrb_\mrj - \sum_\mrj\,\mra_\mrj$. For $\mrz < 0$ the integral
converges if $\uppsi_0 > 0$; when $\uppsi_0 < 0$ we start again from the
original integral and use contiguity relations before introducing the
corresponding MB representation. Introducing the notation
$\hyptt(\mathbf{a}\,;\,\mathbf{b}\,;,\mrz\,;\,\uppsi_0)$
we can use the following contiguity relation where
\bq
\hyptt(\uppsi) = \hyptt(\mathbf{a}\,;\,\mathbf{b}\,;\,\mrz\,;\,\uppsi) \spc \quad
\hyptt(\mra_1 + 1\,;\,\uppsi) =
\hyptt(\mra_1 + 1\,,\,\mra_2\,,\,\mra_3\,;\,\mathbf{b}\,;,\mrz\,;\,\uppsi) \quad \hbox{etc} \spp
\eq
It follows that
\bqa
\mra_1\,(1 - \mrz)\,\hyptt(\mra_1 + 1\,;\,\uppsi_0 - 1) &=&
  \Bigl[ \mra_1 + ( \mra_2 + \mra_3 - \mrb_1 - \mrb_2)\,\mrz\Bigr]\,\hyptt(\uppsi_0)
\nl
{}&+& 
\frac{\prod_{\mrj=1}^{3}\,\eG{\mrb_1 - \mra_\mrj}}{\mrb_1\,(\mrb_1 - \mrb_2)}\,
\ghyp{3}{2}(\mrb_1 + 1\,;\,\uppsi_0 + 1) +
\frac{\prod_{\mrj=1}^{3}\,\eG{\mrb_2 - \mra_\mrj}}{\mrb_2\,(\mrb_2 - \mrb_1)}\,
\ghyp{3}{2}(\mrb_2 + 1\,;\,\uppsi_0 + 1) \spp
\eqa
This relation can be used recursively until all functions have a positive
$\uppsi_0$. We can use the MB representation with $\mrL =
\mrL_{\mri\,\infty}$
but this will return the wrong imaginary part when $0 < \mrz < 1$. The
crucial step in understanding the relations among MB integrals defined on
different contours is based on the transformation $\mrz \to \mrz/(\mrz -
1)$.
There is a simple relation whe we are dealing with a $\ghyp{2}{1}$ function; the
same is not true for $\mrp > 1$. We will discuss the solution by
considering two different strategies. The first one uses a relation
between $\ghyp{3}{2}$ and  $\ghyp{2}{1}$:
\bq
 \hyptt(\mathbf{a}\,;\,\mathbf{b}\,;\,\mrz) =
 \frac{1}{\mrB(\mrb_2 - \mra_3\,,\,\mra_3)}\,
 \int_0^1 \mrd \mrx \,\mrx^{\mra_3 - 1}\,(1 - \mrx)^{\mrb_2 - \mra_3 - 1}\,
 \ghyp{2}{1}(\mra_1\,,\mra_2\,;\,\mrb_1\,;\,\mrz\,\mrx) \spc
\eq
requiring $\Re \mrb_2 > \Re \mra_3 > 0$. Next we transform $\ghyp{2}{1}$,
\bq
\ghyp{2}{1}(\mra_1\,,\,\mra_2\,;\,\mrb_1\,;\,\mrz\,\mrx) =
(1 - \mrz\,\mrx)^{ - \mra_1}\,
\ghyp{2}{1}(\mra_1\,,\,\mrb_1 - \mra_2\,;\,\mrb_1\,;\,\zeta) \spc
\eq
with $\zeta = \mrz\,\mrx/(\mrz\,\mrx - 1) < 0$ for $0 < \mrz < 1$. Therefore we obtain
\bqa\ghyp{3}{2}(\mathbf{a}\,;\,\mathbf{b}\,;\,\mrz) &=&
\frac{1}{\eB{\mra_3}{\mrb_2 - \mra_3}}\,\frac{\eG{\mrb_1}}{\eG{\mra_1}\,\eG{\mrb_1 - \mra_2}}\,
\int_0^1 \mrd \mrx\,\mrx^{\mra_3-1}\,(1 - \mrx)^{\mrb_2-\mra_3-1}\,(1 - \mrz\,\mrx)^{-\mra_1}
\nl
{}&\times&
\int_{\mrL_{\mri\,\infty}}\,\frac{\mrd \mrs_1}{2\,\mri\,\pi}\,\eG{ - \mrs_1}\,
\frac{\eG{\mra_1 + \mrs_1}\,\eG{\mrb_1 - \mra_2 + \mrs_1}}{\eG{\mrb_1 + \mrs_1}}\,
\Bigl( \frac{\mrz\,\mrx}{1 - \mrz\,\mrx} \Bigr)^{\mrs_1} \spp
\eqa
The $\mrx$ integral is
\bqa
\mrJ &=& \int_0^1 \mrd \mrx\,\mrx^{\mra_3 + \mrs_1 - 1}\,(1 - \mrx)^{\mrb_2 - \mra_3 - 1}\,
(1 - \mrz\,\mrx)^{- \mra_1 - \mrs}
\nl
{}&=&
\frac{\eG{\mra_3 + \mrs_1}\,\eG{\mrb_2 - \mra_3}}{\eG{\mrb_2 + \mrs_1}}\,
\ghyp{2}{1}(\mra_1 + \mrs_1\,,\,\mra_3 + \mrs_1\,;\,\mrb_2 + \mrs_1\,;\,\mrz) \spp
\eqa
Finally we transform again $\mrz \to \mrz/(\mrz - 1)$ in the $\ghyp{2}{1}$ function and introduce the
corresponding MB representation. 
In this way we have obtained a MB representation defined along $\mrL_{\mri\,\infty}$ contours when
$0 < \mrz < 1$. The result is
\bqa
\ghyp{3}{2}(\mathbf{a}\,;\,\mathbf{b}\,;\,\mrz) &=&
\frac{          \eG{\mrb_1}
         \,\eG{\mrb_2}}
     {    \eG{\mra_1}
         \,\eG{\mra_3}
         \,\eG{\mrb_1 - \mra_2}
         \,\eG{\mrb_2 - \mra_3}}
\nl
&\times&
\Bigl[ \prod_{\mrj=1}^{2}\,\int_{\mrL_\mrj}\,\frac{\mrd \mrs_\mrj}{2\,\mri\,\pi} \Bigr]\,
         \,\eG{ - \mrs_1}
         \,\eG{ - \mrs_2}\,
\frac{         \eG{\mra_3 + \mrs_1}
         \,\eG{\mra_1 + \mrs_2 + \mrs_1}
         \,\eG{\mrb_1 - \mra_2 + \mrs_1}
         \,\eG{\mrb_2 - \mra_3 + \mrs_2} }
{         \eG{\mrb_1 + \mrs_1}
         \,\eG{\mrb_2 + \mrs_2 + \mrs_1}}
\nl
{}&\times&
\mrz^{\mrs_1 + \mrs_2}\,(1 - \mrz)^{ - \mra_1 - \mrs_1 - \mrs_2} \spp
\eqa
Therefore, using the $\mrp = 2$ example we have shown how to handle the problem of expressing the result
in terms of Fox functions defined on $\mrL_{\mri\,\infty}$, providing the correct imaginary part.

A second approach is bases on the N\o{}rlund identity~\cite{Ncoef,_etinkaya_2021}:
\bq
\frac{\eG{\mathbf{a}_{12}}}{\eG{\mathbf{b}}}\,
\ghyp{\mrp + 1}{\mrp}(\mathbf{a}\,;\,\mathbf{b}\,;\,\mrz) =
\sum_{\mrn=0}^{\infty}\,
\frac{\mrg^{\mrp}_{\mrn}(\mathbf{a}_{12}\,;\,\mathbf{b})}{\eG{\nu_{12} + \mrn}}\,
\ghyp{2}{1}(\mra_1\,,\,\mra_2\,;\,\nu_{12} + \mrn\,;\,\mrz) \spc
\eq
where we have defined 
\bq
\mathbf(a) = (\mra_1\,\dots\,\mra_{\mrp+1}) \spc \quad
\eG{\mathbf{a}} = \prod_{\mrj}\,\eG{\mra_\mrj} \spc \quad
\nu_{12} = \sum_{\mrj=1}^{\mrp}\,\mrb_\mrj - \sum_{\mrj=3}^{\mrp}\,\mra_\mrj \spc \quad
\mathbf{a}_{12} = (\mra_3\,\dots\,\mra_{\mrp+1}) \spp
\eq
For $\mrp = 2$ we have $\nu_{12} = \mrb_1 + \mrb_2$, $\mathbf{a}_{12} = \mra_3$ and
\bq
\mrg^{2}_{\mrn} = \frac{1}{\mrn\,!}\,(\mrb_2 - \mra_3)_{\mrn}\,(\mrb_1 - \mra_3)_{\mrn} \spp
\eq
Performing the transformation $\mrz \to \mrz/(\mrz - 1)$ in $\ghyp{2}{1}$, followed by introducing the
corresponding MB representation, we can write $\ghyp{3}{2}$ as a series of $\mrH^{2\,,\,1}_{2\,,\,2}$ Fox
functions of argument $\mrz/(1 - \mrz)$.
\paragraph{Example $4$} \hspace{0pt} \\
Consider the following integral:
\bq
 \mrI = \int_0^1 \mrd \mrx\,\mrx^{\mra - 1}\,(1 - \mrc)^{\mrc - \mra -
1}\,\prod_{\mrj=1}^{2}\,
 ( 1 - \mrz_\mrj\,\mrx)^{ - \mrb_\mrj} \spp
\eq
For the sake of simplicity let us assume $\mrb_1 = \mrb_2 = 1$; therefore we have
\bq
  \prod_{\mrj=1}^{2}\,(1 - \mrz_\mrj\,\mrx)^{-1} =
  \frac{1}{\mrx}\,\frac{1}{\mrz_1 - \mrz_2}\,
  \Bigl( \frac{1}{1 - \mrz_1\,\mrx} - \frac{1}{1 - \mrz_2\,\mrx} \Bigr) \spp
\eq
The integral is a Lauricella function $\mrF^{(2)}_{\mrD}$ but we want to
proceed by using
\bq
(1 - \mrz_\mrj\,\mrx)^{-1} = \int_{\mrL}\,\frac{\mrd
\mrs}{2\,\mri\,\pi}\,\eB{ - \mrs}{1 + \mrs}\,( -
\mrz_\mrj\,\mrx)^{\mrs} \spc
\eq
where $\mrL$ can be $\mrL_{\pm\,\infty}$ depending on $\mrz_{\mrj}$. After
performing the $\mrx$ integral we obtain $\mrI = \mrI_1 - \mrI_2$ with
\bq
 \mrI_\mrj = - \frac{\Gamma(\mrc - \mra)}{\mrz_1 -
\mrz_2}\,\int_{\mrL}\,\frac{\mrd \mrs}{2\,\mri\,\pi}\,\Gamma( -
\mrs)\,\frac{\Gamma(1 + \mrs)\,\Gamma(\mra - 1 + \mrs)}{\Gamma(\mrc -1 +
\mrs)}\,( - \mrz_\mrj)^{\mrs} \spp
\eq
According to the notations of \Bref{compH} the MB integral corresponds to
parameters $\mrm = 2, \mrn = 1$ and $\mrp = \mrq = 2$ giving $\mu = 0$ and
$\beta = 1$. As a consequence, $\mid \mrz_\mrj \mid > 1$ corresponds to
$\mrL = \mrL_{-\,\infty}$ while $\mid \mrz_\mrj \mid < 1$ corresponds to
$\mrL = \mrL_{+\,\infty}$. We have A{-}poles at $ \mrs = \mrm_1$ and B{-}
poles at $\mrs = -1 - \mrm_2, \mrs = 1 - \mra - \mrm_3$.
The result is
\bq
\mrI_{\mrj\,\pm} = \frac{\eG{\mrc - \mra}}{\mrz_1 - \mrz_2}\,\sum_{\mrm=0}^{\infty}\,
\mrJ_{\mrj\,,\,\pm\,,\,\mrm} \spc
\eq
\bqa
\mrJ_{\mrj\,,\,+\,,\,\mrm} &=&
\frac{\eG{\mra - 1 + \mrm}}{\eG{\mrc - 1 + \mrm}}\,\mrz_\mrj^{\mrm} \spc
\nl
\mrJ_{\mrj\,,\,-\,,\,\mrm} &=&
\frac{\eG{\mra - 2 - \mrm}}{\eG{\mrc - 2 - \mrm}}\,( - \mrz_{\mrj})^{ - \mrm - 1} +
\frac{\eG{\mra - 1 + \mrm}\,\eG{2 - \mra - \mrm}}{\eG{1 + \mrm}\,\eG{\mrc - \mra - \mrm}}\,
( - \mrz_{\mrj})^{1 - \mra - \mrm} \spp
\eqa
Obviously we can wrire
\bq
\mrI = \frac{\eG{\mra}\,\eG{\mrc - \mra}}{\eG{\mrc}}\,
\mrF^{(2)}_{\mrD}(\mra\,;\,\mathbf{b}\,;\,\mrc\,;,\mrz_1\,,\,\mrz_2) \spc
\label{IFD}
\eq
and we would like to understand when we can use
\bq
\mrF^{(2)}_{\mrD} = \frac{\eG{\mrc}}{\eG{\mra}\,\eG{\mrb_1}\,\eG{\mrb_2}}\,
\Bigl[ \prod_{\mrj=1}^{2}\,\int_{\mrL_\mrj}\,\frac{\mrd \mrs_\mrj}{2\,\mri\,\pi} \Bigr] \,
\frac{\eG{\mra + \mrs_1 + \mrs_2}}{\eG{\mrc + \mrs_1 + \mrs_2}}\,
\prod_{\mrj=1}^{2}\,\eG{ - \mrs_\mrj}\,\eG{\mrb_\mrj + \mrs_\mrj}\,( - \mrz_\mrj)^{\mrs_\mrj} \spc
\label{IMBt}
\eq
with $\mrL_\mrj = \mrL_{\mri\,\infty}$ and when $\mrz_1$ and/or $\mrz_2$ are real and positive.
The integral in \eqn{IMBt} can be convergent (depending on $\mra$ and $\mrc$) but the question is about the imaginary part.
Comparing $\mrL_{\pm}$ with $\mrL_{\mri\,\infty}$ shows the expected result:
\bq
\Re\,\mrI_{\mri\,\infty} = \Re\,\mrI_{\pm} \spc \qquad
\Im \mrI_{\mri\,\infty} = \Im\,\mrI_{\pm} \spc \quad \hbox{iff} \qquad \mrz_{1,2} > 1 \spp
\eq
The solution is as follows: using \eqn{IMB} there are two options:
\begin{itemize}

\item $0 < \mrz_\mrj < 1$. We transform

\bq
\mrF^{(2)}_{\mrD}(\mra\,;\,\mathbf{b}\,;\,\mrc\,;\,\mathbf{z}) =
\Bigl[ \prod_{\mrj=1}^{2}\,(1 - \mrz_\mrj)^{ - \mrb_\mrj} \Bigr]\,
\mrF^{(2)}_{\mrD}(\mrc - \mra\,;\,\mathbf{b}\,;\,\mrc\,;\,\zeta_1\,,\,\zeta_2) \spc
\eq
with $\zeta_{\mrj} = \mrz_\mrj/(\mrz_\mrj - 1)$. Since $\zeta_\mrj < 0$ we can use the corresponding MB representation.

\item $0 < \mrz_1 < 1$ and $\mrz_2 > 1$. We transform

\bq
\mrF^{(2)}_{\mrD}(\mra\,;\,\mathbf{b}\,;\,\mrc\,;\,\mathbf{z}) =
(1 - \mrz_1)^{ - \mra}\,\mrF^{(2)}_{\mrD}(\mra\,;\,\mrc - \mrb_1 - \mrb_2\,,\,\mrb_2\,;\,\mrc\,;\,\zeta_1\,,\,\zeta_2) \spc
\eq
where $\zeta_1 = \mrz_1/(\mrz_1 - 1)$ and $\zeta_2 = (\mrz_2 - \mrz_1)/(1 - \mrz_1)$. Since
$\zeta_1 < 0$ and $\zeta_2 > 1$ we can use the corresponding MB representation.  

\end{itemize}

The conclusion is as follows: when computing a Feynman integral in the Feynman parametrizatio, compute the first
integral producing a Lauricella function; transform it (if needed) before using the MB representation; use the
MB representation to compute the second integral and repeat the procedure.

As an additional example consider the following integral:
\bq
\mrH = \int_0^1 \mrd \mrx\,\mrd \mry\,\mrx^{\mra - 1}\,(1 - \mrx)^{\mrc - \mra - 1}\,\,
(\mrx^2 - 2\,\lambda\,\mrx - \mry^2)^{ - \mrb} \spp
\eq
We have roots
\bq
\mrx_{\pm} = \lambda \pm \sqrt{\lambda^2 + \mry^2} \spc \qquad
\lambda = \lambda - \mri\,\delta \spc \quad \delta \to 0_{+} \spp
\eq
With $\lambda > 1$ we obtain $\mrx_{+} > 1$ and $\mrx_{-} < 0$ and
\bq
\mrH = \int_0^1 \mrd \mrx\,\mrd \mry\,\mrx^{\mra - 1}\,(1 - \mrx)^{\mrc - \mra - 1}\,
(\mrx - \mrx_{-})^{ - \mrb}\,(\mrx - \mrx_{+})^{- \mrb} \spp
\eq
The $\mrx$ integration gives a Lauricella function of arguments 
$1/\mrx_{-} < 0$ and $0 < 1/\mrx_{+} < 1$.
We use the transformation
\bq
\mrF^{(2)}_{\mrD}(\mra\,;\,\mathbf{b}\,;\,\mrc\,;\,\mathbf{z}) =
(1 - \mrz_1)^{\mrc - \mra - \mrb_1}\,(1 - \mrz_2)^{\mrb_2}\,
\mrF^{(2)}_{\mrD}(\mrc - \mra\,;\,\mrc - \mrb_1 - \mrb_2\,,\,\mrb_2\,;\,\zeta_1\,,\,\zeta_2) \spc
\eq
\bq
\zeta_1 = \mrz_1 = 1/\mrx_{-} \spc \quad 
\mrz_2 = 1/\mrx_{+} \spc \quad \zeta_2 = \frac{\mrx_{+} - \mrx_{-}}{\mrx_{-}\,(\mrx_{+} - 1)} \spp
\eq
Since $\zeta_\mrj < 0$ we can use the MB representation obtaining
\bqa
\mrH &=& \frac{\eG{\mra}}{\eG{\mrc - 2}}\,\int_0^1 \mrd \mry\,
(\lambda - \eta)^{\mra - \mrc}\,(\lambda - \eta - 1)^{\mrc - \mra - 1}
\nl
{}&\times&
\Bigl[ \prod_{\mrj=1}^{2}\,\frac{\mrd \mrs_\mrj}{2\,\mri\,\pi}\,
\frac{\eG{\mrc - \mra + \mrs_1 + \mrs_2}}{\eG{\mrc + \mrs_1 + \mrs_2}}\,
\eG{ - \mrs_1}\,\eG{\mrc - 2 + \mrs_1}\,\eG{ - \mrs_2}\,\eG{1 + \mrs_2}
\nl
{}&\times&
(2\,\eta)^{\mrs_2}\,(\eta - \lambda)^{ - \mrs_1 - \mrs_2}\,(\lambda + \eta - 1)^{- 1 + \mrs_2} \spc
\eqa
where we have introduced $\eta^2 = \lambda^2 + \mry^2$. We can continue to integate using
\bqa
\mry &=& 2\,\lambda\,\frac{\mrt}{\mrt^2 - 1} \spc \qquad
\mrd \mry = 2\,\lambda\,\frac{\mrt^2 + 1}{(\mrt^2 - 1)^2}\,\mrd \mrt \spc
\nl
\eta &=& \lambda\,\frac{\mrt^2 + 1}{\mrt^2 - 1} \spc \qquad
\mrt_ 0 \le \mrt \le \infty \spc \quad \mrt_0 = \frac{1}{\sqrt{1 + \lambda^2} - \lambda} > 1 \spp
\eqa
Therefore the last integration can be performed giving again Lauricella functions. 
\paragraph{Complex masses} \hspace{0pt} \\
We will not discuss general aspects concerning the definition and the use
of this scheme, for details see \Brefs{Actis:2006rc,Passarino:2010qk,Passarino:2018wix}.
We simply assume that internal masses in Feynman integrals are replaced by
complex poles $\mrm^2 \to \mu^2 - \mri\,\gamma\,\mu$. Contrary to the
naive expectation the use of complex poles does not help in the use of MB
splitting. This is due to the fact that complex poles do not lie on the
first Riemann sheet. Let us consider a simple example:
\bq
\mrI = \int_0^1\,\mrd \mrx\,\bigl(\mrs\,\mrx^2 - \mrs\,\mrx + \mrm^2
\bigr)^{-1} = - \frac{1}{\eta\,\mrs} \,
\int_0^1\,\mrd \mrx\,\bigl[ \mrx\,(1 - \mrx) - \eta \bigr]^{-1} \spc
\eq
where $\eta$ is defined by
\bq
\eta = \frac{\mu^2}{\mrs} - \,\mri\,\frac{\gamma\,\mu}{\mrs} , \qquad \mrs > 0 \spp
\eq
We obtain
\bq
\mrI = -\,\frac{1}{\eta\,\mrs}\,
\int_{\mrL}\,\frac{\mrd \mrr}{2\,\mri\,\pi}\,
\Gamma^3(\mrr)\,\frac{\Gamma(1 - \mrr)}{\Gamma(2\,\mrr)}\,( - \eta )^{- \mrr} \spp
\eq
If we define $\phi = \pi - \arctan(\gamma/\mu)$ we obtain
\bq
( - \eta )^{ - \mrr} = \exp\{- \mrr\,\xi\} , \qquad
\xi = \ln\,\mid \eta \mid + \,\mri\,(\pi - \phi) - 2\,\mri\,\pi \spc
\eq
where we have taken the logarithm on the second Riemann sheet. Writing
$\mrr = \sigma + \mri\,\mrt$ we get
\bq
( - \eta )^{- \mrr} = \mid \eta \mid^{- \sigma}\,
\exp\{\rho(\mrt)\} \spc
\eq
where $\rho$ is defined by
\bq
\rho = - (\pi + \phi)\,\mrt + \mri\,\Bigl[
\sigma\,(\pi + \phi) - \mrt\,\ln \mid \eta \mid \Bigr] \spp
\eq
The MB integral corresponds to parameters $\alpha = 2$,
$\beta = 0$ and $\lambda = 1/2$. The absolute value of the integrand is
compatible with
\bq
\mid \mrt \mid^{1/2}\,\exp\{ - \pi\,\mid \mrt \mid - (\pi + \phi)\,\mrt\} \spc
\eq
therefore the MB integral does not converge when $\mrL = \mrL_{\mri\,\infty}$.
In conclusions we have considered cases where the answer is given by Fox functions (usually Lauricella functions)
depending on variables $\mrz_1\,\dots\,\mrz_\mrn$. The crucial point consists in the relation between the
Euler{-}Mellin and the Mellin{-}Barnes representations. There are different sectors, corresponding to
$0 < \mrz_\mrj < 1$ or $\mrz_\mrj > 1$; we have provided the set of transformations needed to have MB representations, 
defined on $\mrL_{\mri\,\infty}$, which reproduce the correct cut structure of the original integral.
\section{Feynman integrals \label{exa}}
Before introducing specific examples we point out an important relation:
\bq
\lpar \mra\,\mrb \rpar^{\alpha} = \mra^{\alpha}\,\mrb^{\alpha}\,\exp\{ \alpha\,\eta(\mra\,,\,\mrb) \} \spc \quad
\eta(\mra\,,\,\mrb) = \tip\,\Bigl[
\theta( - \Im \mra)\,\theta( - \Im \mrb)\,\theta( \Im \mra\,\mrb) -
\theta(\Im \mra)\,\theta(\Im \mrb)\,\theta( - \Im \mra\,\mrb) \Bigr] \spp
\eq
In particular, when $\mra, \mrb \in \Rf$ with $\mra > 0$ and $\mrb < 0$ we have
\bq
\lpar \mra\,\mrb - \mri\,\delta \rpar^{\alpha} = \mra^{\alpha}\,\lpar \mrb - \mri\,\delta \rpar ^{\alpha} \spc \qquad
\delta \to 0_{+} \spp
\eq
\subsection{Virtual corrections \label{virt}}
Examples of Feynman integrals involved in the calculation of virtual corrections are:
\paragraph{Triangle} \hspace{0pt} \\
Consider the following integral corresponding to a three{-}point Feynman
function in arbitrary space{-}time dimensions ($\mrd = 4 + \ep$), with a normal threshold at
$\mrs = 4\,\mrm^2$ ($\lambda = 1/4$)
\bq
\mrC = \pi^{\ep/2}\,\Gamma(1 - \ep/2)\,
\int_0^1 \mrd \mrx\,\int_0^{\mrx} \mrd \mry\,
\Bigl[ \mrm^2\,(1 - \mrx) + (\mrm^2 - \mrs)\,\mry + \mrs\,\mrx\,\mry
\Bigr]^{\ep/2 - 1} \spc
\eq
The integral can be written in terms of
an ${}_3\,\mrF_2$ hypergeometric function.
\bq
\mrC = \pi^{\ep/2}\,\mrm^{\ep - 2}\,
\frac{\Gamma^2(1 - \ep/2)\,\eG{1 + \ep/2}}{\eG{2 + \ep/2}}\,
{}_3\mrF_2(1\,,\,1 - \ep/2\,,\,1 + \ep/2\,;\,\frac{3}{2}\,,\,2 +
\ep/2\,;\,\frac{1}{4\,\lambda} ) \spc
\eq
with $\lambda= \mrm^2/\mrs$.
For the generalized hypergeometric function we can write the following
representation:
\bq
{}_3\,\mrF_2(\mathbf{a}\,;\,\mathbf{b}\,;\,\mrz) =
\frac{\Gamma(\mrb_1)\,\Gamma(\mrb_2)}{\Gamma(\mra_3)}\,
\int_{\mrL}\,\frac{\mrd \mrs}{2\,\mri\,\pi}\,
\frac{\mrA(\mrs)}{\Gamma(\mrc + \mra_1 + \mra_2 + \mrs)}\,
{}_2\,\mrF_1(\mra_1\,,\,\mra_2\,;\,\mrc + \mra_1 + \mra_2 + \mrs\,;,\,\mrz) \spc
\eq
\bq
\mrA(\mrs) = e^{\mri\,\pi\,\mrs}\,\Gamma( - \mrs)\,
\frac{
\Gamma(\mrc + \mra_1 + \mra_2 - \mrb_1 + \mrs)\,
\Gamma(\mrb_1 - \mra_3 + \mrs)}
{\Gamma(\mrb_1 - \mra_3)\,\Gamma(\mrb_2 - \mra_3)} \spp
\eq
For the Gauss hypergeometric function of argument $\mrz =
(4\,\lambda)^{-1}$ we can use previous results to write a MB
representation valid for $0 < s < 4\,\mrm^2$ and for $\mrs > 4\,\mrm^2$.

Now we set $\ep = 0$ and study the behavior of $\mrC$ below the normal threshold; we
start with the region $0 < \mrs < \mrm^2$, where $\lambda > 1$. We obtain
\bq
\mrC = \frac{1}{\mrs}\,\int_0^1\,\mrd \mrx\,\frac{\mrX}{\lambda - \mrx}\,
\hyp{1}{1}{2}{ - \mrX} \spc
\qquad
\mrX = \frac{(\lambda - \mrx)\,(1 - \mrx)}{\lambda\,\mrx} \spp
\eq
Since $\mrX > 0$ we can use a MB representation for the hypergeometric function and perform the $\mrx$ integration
obtaining a new hypergeometric function of argument $1/\lambda > 0$; therefore we use
\bq
\hyp{\mra}{\mrb}{\mrc}{\lambda^{-1}} =
\lpar 1 - \frac{1}{\lambda} \rpar^{ - \mra}\,\hyp{\mra}{\mrc - \mrb}{\mrc}{\frac{1}{1 - \lambda}} \spp
\eq
Using again the corresponding MB representation we get
\bq
\mrC = \int_{\mrL_1}\,\frac{\mrd \mrs_1}{\tip}\,\eGs{ - \mrs_1}\,\eGs{1 + \mrs_1}\,
\lpar 1 - \frac{1}{\lambda} \rpar^{\mrs_1}\,
\hyp{ - \mrs_1}{2 + \mrs_1}{2}{\frac{1}{1 - \lambda}} \spc
\eq
and we can use the standard MB representation for the hypergeometric
function and no imaginary part will arise.

Next we consider the region $\mrm^2 < \mrs < 4\,\mrm^2$ where $1/4 < \lambda < 1$. In this case it is
more convenient to split the $\mrx$ integration introducing $\mrC_{<}$ where $0 < \mrx < \lambda$ and
$\mrC_{>}$ where $\lambda < \mrx < 1$.
We obtain
\bq
\mrC_{<} = \frac{1}{s}\,\int_{\mrL_1}\,\frac{\mrd \mrs_1}{\tip}\,
\frac{\eGs{ - \mrs_1}\,\eGc{1 + \mrs_1}}{\eG{2 + \mrs_1}}\,
\lambda^{-1-\mrs_1}\,(1 - \lambda)^{1+\mrs_1}\,
\hyp{-1 - \mrs_1}{1 + \mrs_1}{1}{- \frac{\lambda}{1 - \lambda}} \spp
\eq
Also in this case we can use the standard MB representation for the hypergeometric
function and no imaginary part will arise.

The $\mrC_{>}$ integral can be rewritten as
\bq
\mrC_{>} = \frac{1}{s}\,\int_0^1\,\frac{\mrd \mrx}{\mrx } \,\mrX\,
{}_2\mrF_1(1\,,\,1\,;\,2\,;\,\mrX)
\spc \qquad
\mrX = \frac{(1 - \lambda)^2}{\lambda}\,
\frac{\mrx\,(1 - \mrx)}{(1 - \lambda)\,\mrx  + \lambda} \spp
\eq
In this way we obtain
\bq
 \mrx_{\pm} = \frac{1}{2}\,(1 - \lambda)^{-1}\,
 \Bigl[1 - 2\,\lambda \pm \mri\,\sqrt{4\,\lambda - 1} \Bigr] \spc \qquad
 \frac{\mrX}{\mrX - 1} = -\,
 \frac{\mrx\,(1 - \mrx)}{(\mrx - \mrx_-)\,(\mrx - \mrx_+)} < 0 \spc
 \eq
with $\mrx_{\pm}$ complex. Therefore we obtain
\bq
 \mrC_{>} = \frac{1}{\mrs}\,
 \int_{\mrL_1}\,\frac{\mrd \mrs_1}{2\,\mri\,\pi}\,
 \Gamma( - \mrs_1)\,
 \frac{\Gamma^2(1 + \mrs_1)}{\Gamma(2 + \mrs_1)}\,\mrJ \spc
\eq
where $\mrJ$ is
\bq
 \mrJ = \bigl( \frac{\lambda}{1 - \lambda} \bigr)^{2\,\mrs_1}\,
 \frac{\Gamma(1 + \mrs_1)\,\Gamma(2 + \mrs_1)}{
 \Gamma(3 + 2\,\mrs_1)}\,
 \mrF^{(2)}_{\mrD}\bigl(
 1 + \mrs_1\,;\,1 + \mrs_1\,,\,1 + \mrs_1\,;\,3 +
2\,\mrs_1\,;\,\frac{1}{\mrx_-}\,,\,\frac{1}{\mrx_+} \bigr) \spp
\eq
Since $\mrx_{\pm}$ are complex we can use the standard MB represntation
for the Lauricella function.
\paragraph{Box} \hspace{0pt} \\
Next we consider the following integral:
\bq
\mrD = \int_0^1 \mrd \mrx\,\int_0^{\mrx} \mrd \mry\,\int_0^{\mry} \mrd \mrz\,
\Bigl[
( \mru\,\mrx + \mrs\,\mry + \mrm^2 + \mrt)\,\mrz +
\mrt\,\mrx\,\mry + \mrm^2\,\mrx - (\mrm^2 + \mrt)\,\mry
]^{-2} \spc
\eq
which corresponds to a four{-}point function with a normal threshold at
$\mrs = 4\,\mrm^2$. After performing the $\mrz$ integration followed by
the $\mrx$ integration we obtain $\mrD = \sum_{\mrj=1}^{3}\,\mrD_{\mrj}$,
\bq
\mrD_1 = - \int_0^1\,\frac{\mrd \mrx}{\mrQ_1}\,
\ln(1 + \mrX_1) \spc \quad
 \mrD_2 = \int_0^1 \mrd \mrx\,\Bigl(
 \frac{1}{\mrQ_1} - \frac{1}{\mrQ_2} \bigr)\,
 \ln(1 + \mrX_2) \spc \quad
 \mrD_3 = \int_0^1 \frac{\mrd \mrx}{\mrQ_2}\,
 \ln(1 + \mrX_3) \spp
\eq
The two quadratic forms are:
\bq
\mrQ_1 = - \mrs\,\mrt\,(\mrx - \mrx_-)\,(\mrx - \mrx_+) , \qquad
\mrQ_2 = - \mrs\,\mrt\,(\mrx^2 - \mrx_0) \spc
\eq
where we have introduced
\bq
 \mrx_0 = \frac{\mrm^2}{\mrs\,\mrt}\,(\mru - \mrm^2) \spc \quad
 \mrx_{\pm} = - \frac{1}{2\,\mrt}\,\bigl( \mru \mp \sqrt{\Delta} \bigr) ,
\quad
 \Delta = \mru^2 - 4\,\mrm^2\,\frac{\mrt}{\mrs}\,(\mrm^2 - \mru) \spp
\eq
Having introduced $\lambda = \mrm^2/\mrs$ we have
\bq
 \mrX_1 = \frac{\mrx - \lambda}{\lambda - \mrx + \mrx^2}\,(1 - \mrx) \spc \quad
 \mrX_2 = - \frac{\mru}{\mrm^2 + \mrs\,\mrx}\,(1 - \mrx) \spc \quad
 \mrX_3 = - 1 - \frac{\mrt}{\mrm^2}\,\mrx \spp
\eq
With $\mru = - \mrs - \mrt$ and $- \mrs < \mrt < 0$
we examine the region $0 < \mrs < \mrm^2$. In this region $\lambda - \mrx
+ \mrx^2 > 0$; furthermore, $\lambda > 1$ which means that
$\mrx < \lambda$ for $\mrx \in [0\,,\,1]$. It
follows that $\mrX_1 < 0$.
In the same region we have $\mrX_2 > 0$ and $\mrX_3 < 0$.
The strategy is to write
\bq
\ln(1 + \mrX) = \mrX\,
{}_2\mrF_1(1\,,\,1\,;\,2\,;\, - \mrX)
\eq
and use the MB representation for the hypergeometric function, eventually
requiring a transformation in order to respect the conditions which are
needed. The simplest integral is $\mrD_3$ where $\mrX_3 < 0$ and
$\mrX_3/(1 + \mrX_3) < 0$. We introduce
\bq
 \mrR(\mrs) = \Gamma( - \mrs)\,
 \frac{\Gamma^2(1 + \mrs)}{\Gamma(2 + \mrs)} \spc
\eq
and write
\bq
 \mrD_3 = \int_{\mrL_1}\,\frac{\mrd
\mrs_1}{2\,\mri\,\pi}\,\mrR(\mrs_1)\,\mrJ_3 \spc \quad
 \mrJ_3 = - \frac{1}{2\,\mrm^2\,(\mru - \mrm^2)}\,
 \bigl( - \frac{\mrt}{\mrm^2} \bigr)^{- 1 - \mrs_1}\,( \mrJ_{3\,-} -
\mrJ_{3\,+} ) \spc
\eq
\bq
\mrJ_{3\,-} = \frac{\Gamma( - \mrs_1)}{\Gamma(1 - \mrs_1)}\,
\mrF^{(2)}_{\mrD}\bigl(
- \mrs_1\,;\,- 1 - \mrs_1\,,\,1\,;\,1 - \mrs_1\,;\, -
\frac{\mrt}{\mrm^2}\,,\,\frac{1}{\mrx_0}
 \bigr) \spp
 \eq
 The arguments of the Lauricella function are
 $\mrz_1 = - \mrt/\mrm^2$ and $\mrz_2 = 1/\mrx_0$
 and are both positive. Therefore we first transform the Lauricella
function and obtain
 \bq
  \mrF^{(2)}_{\mrD}\bigl(
- \mrs_1\,;\,- 1 - \mrs_1\,,\,1\,;\,1 - \mrs_1\,;\, -
\frac{\mrt}{\mrm^2}\,,\,\frac{1}{\mrx_0}
 \bigr) =
 \bigl( 1 + \frac{\mrt}{\mrm^2} \bigr)^{1 + \mrs_1}\,\bigl(1 -
\frac{1}{\mrx_0} \bigr)^{-1}\,
 \mrF^{(2)}_{\mrD}\bigl(
 1\,;\,- 1 - \mrs_1\,,\,1\,;\,1 - \mrs_1\,;\,
 \frac{\mrt}{\mrt + \mrm^2}\,,\,\frac{1}{1 - \mrx_0} ] \bigr) \spc
 \eq
and we can use the standard MB representation of the Lauricella function.
The remaining integral is
\bq
\mrJ_{3\,+} = \frac{\Gamma( - \mrs_1)}{\Gamma(1 - \mrs_1)}\,
\mrF^{(2)}_{\mrD}\bigl(
- \mrs_1\,;\,- 1 - \mrs_1\,,\,1\,;\,1 - \mrs_1\,;\, -
\frac{\mrt}{\mrm^2}\,,\, -\frac{1}{\mrx_0} \spc
 \bigr)
 \eq
requiring a transformation of the Lauricella function before
we can use the standard MB representation.

The other quadratic is $\mrQ_1$ with roots $\mrx_{\pm}$ and (for $0 < \mrs
< \mrm^2$)
$\mrx_- < 0$, $\mrx_+ > 1$, so that
\bq
\mrD_1 = - \int_{\mrL_1}\,\frac{\mrd \mrs_1}{2\,\mri\,\pi}\,
\mrR(\mrs_1)\,\mrJ_1
\spc \quad
  \mrJ_1 = \frac{1}{\mrs\,\mrt}\,
  \int_0^1\,\frac{\mrd \mrx}{(\mrx - \mrx_-)\,(\mrx - \mrx_+)}\,\bigl(
  \frac{\lambda\,\mrx}{\lambda - \mrx + \mrx^2} \bigr)^{1 + \mrs_1} \spp
 \eq
Furthermore $\lambda - \mrx + \mrx^2 =(\mrx - \mrX_-)\,(\mrx - \mrX_+)$ with
\bq
 \mrX_{\pm} = \frac{1}{2}\,\bigl( 1 \pm \mri\,\sqrt{4\,\lambda - 1} \bigr)
\eq
with $\mrX_{\pm}$ complex. We obtain
\bq
\mrJ_1 = \frac{\mrt}{\sqrt{\Delta}\,\mrm^2\,(\mrm^2 - \mru)}\,(\mrJ_{1\,+}
- \mrJ_{1\,-}) \spc
\eq
\bq
 \mrJ_{1\,\,\pm} = \frac{\Gamma(2 + \mrs_1)}{\Gamma(3 + \mrs_1)}\,
 \mrF^{(3)}_{\mrD}\bigl(
 2 + \mrs_1\,;\,1\,,\,1 + \mrs_1\,,\,1 + \mrs_1\,;\,3 +
\mrs_1\,;\,\frac{1}{\mrx_{\pm}}\,,\,\frac{1}{\mrX_-}\,,\,\frac{1}{\mrX_+}
\bigr) \spp
\eq
For $\mrJ_{2\,-}$ we can use the standard MB representation since $\mrx_-
< 0$. For $\mrJ_{2\,+}$, taking into account that $\mid \mrX_{\pm} \mid =
\lambda > 1$ we use a transformation of the Lauricella function.
\paragraph{Sunrise} \hspace{0pt} \\
We consider a sunrise integral~\cite{Adams:2016sob,Remiddi:2016gno} with non{-}canonical powers and equal
internal masses,
\bq
[ \mrq_1^2 + \mrm_1^2 ]^2 \spc \qquad
[ (\mrq_1 - \mrq_2 + \mrp)^2 + \mrm_2^2 ]^2 \spc \qquad
[ \mrq_2^2 + \mrm_3^2 ] \spp
\eq
After partial quadratization of the Symanzik polynomials (see Sect.~$6$ of \Bref{Passarino:2024ugq}) we obtain
\bq
 \mrS = \int_0^1\,\mrd \rho\,\mrd \mrx\,\rho^3\,(1 - \mrx)\,(\mra\,\mrx^2
+ \mrb\,\mrx + \mrc)^{-1}
\eq
with parameters
\bq
\mra = - \mrb = \rho\,(\mrm^2 - \mrs\,\sigma) \spc \quad
\mrc = - \mrm^2\,\sigma \spc \quad
\sigma = 1 - \rho \spp
\eq
The integral can be rewritten as
\bq
 \mrS = \frac{1}{\mrm^2}\,
 \int_0^1\,\mrd \rho\,\mrd \mrx\,
 \rho^3\,(1 - \mrx)\,
 \Bigl[\mra\,(\mrx - \frac{1}{2})^2 + \mrB \bigr]^{-1} \spc
\eq
where we have introduced
$\mra = \lambda\,\rho\,(\rho - \rho_0)$, $\rho_0 = 1 - 1/\lambda$ and
 \bq
 \mrB = - \frac{1}{4}\,\lambda\,(\rho - \rho_-)\,(\rho - \rho_+) \spc \qquad
 \rho_{\pm} = \frac{1}{2\,\lambda}\,\Bigl[
 \lambda + 3 \pm\sqrt{(\lambda - 1)\,(\lambda - 9)}
 \Bigr]
 \eq
It follows that $\lambda = 1$ corresponds to the pseudo{-}threshold while
$\lambda = 9$ corresponds to the normal threshold. We have four different
regions:

\begin{enumerate}
\item $\lambda < 0$, where $\rho_0 > 1$ and
    $\rho_- > 1$, $\rho_+ < 0$.
\item $0 < \lambda < 1$, where $\rho_0 < 0$ and $\rho_{\pm} > 1$.
\item $1 < \lambda < 9$, where $0 < \rho_0 < 1$ and $\rho_{\pm}$ are complex
\item $\lambda > 9$, where $0 < \rho_- < \rho_+  < 1$
\end{enumerate}
In all cases we always start with
\bq
\mrS = -\,2\,\int_0^1 \mrd \rho\,\rho^3\,
\Bigl[(\rho - \rho_-)\,(\rho - \rho_+) \Bigr]^{-1}\,
{}_2\mrF_1\bigl(1\,,\,\frac{1}{2}\,;\,\frac{3}{2}\,;\,\frac{\rho\,(\rho -
\rho_0)}{(\rho - \rho_-)\,(\rho - \rho_+)} \bigr) \spp
\eq
The strategy will be as follows:
\begin{itemize}
\item for ${}_2\mrF_1(\dots\,;\,\mrz < 0)$ we use the standard MB
representation.
\item For ${}_2\mrF_1(\dots\,;\,\mrz > 0)$ and
$0 < \mrz < 1$  we transform the hypergeometric function before using the
MB representation.
\end{itemize}
After that we perform the $\rho$ integration, obtaining an
$\mrF^{\mrN)}_{\mrD}$ Lauricella function. If needed we transform the
Lauricella function (below the normal threshold) so that we always have to
deal with Lauricella functions with negative arguments; only at this point
we use the corresponding MB representation. The resulting expressions
below the normal threshold are the following ones:
\bqa
\mrS\lpar \lambda < 0 \rpar &=& 
\frac{1}{\pi\,\lambda}\,\Bigl[ \prod_{\mrj=1}^{2}\,\int_{\mrL_{\mrj}}\,\frac{\mrd \mrs_\mrj}{\tip} \Bigr]\,
\frac{
\eG{ - \mrs_1}\,
\eGs{1/2 + \mrs_1}\,
}
{
\eG{3/2 + \mrs_1}\,
}
\,
\eG{ - \mrs_2}\,
\eG{1/2 + \mrs_2}\,
\eG{ - \mrs_2 - \mrs_1}\,
\eG{4 + \mrs_2 + \mrs_1}
\nl
{}&\times&
\lpar - \frac{\lambda}{4} \rpar^{\mrs_1}\,
\lpar \rho_0 - 1 \rpar^{\mrs_1}
\lpar 1 - \frac{1}{\rho_+} \rpar^{\mrs_1}
\lpar 1 - \frac{1}{\rho_-} \rpar^{ - 1/2}
\nl
{}&\times&
\mrF^{(2)}_{\mrD}\lpar - \mrs_1 - \mrs_2\,;\,\frac{1}{2}\,,\, - \mrs_1\,;\,4\,;\,
\frac{1}{1 - \rho_-}\,,\,\frac{1}{1 - \rho_0} \rpar \spp
\eqa
Since $\rho_ 0 > 1$ and $\rho_- > 1$ we can use the standard MB representation for $\mrF^{(2)}_{\mrD}$ and
no imaginary part will emerge.

\bqa
\mrS\lpar 0 < \lambda < 1 \rpar &=& - \frac{1}{\sqrt{\pi}}\,\Bigl[ \prod_{\mrj=1}^{2}\,
\int_{\mrL_\mrj}\,\frac{\mrd \mrs_\mrj}{\tip} \Bigr]\,
\frac{
\eGs{1/2 + \mrs_1}\,
}
{
\eG{3/2 + \mrs_1}\,
\eG{9/2 + \mrs_1}\,
} \,
\eG{ - \mrs_2}\,
\eG{1/2 - \mrs_2}\,
\eG{\mrs_2 - \mrs_1}\,
\eG{4 + \mrs_2 + \mrs_1}
\nl
{}&\times&
\lpar 1 - \rho_0 \rpar^{\mrs_1}\,
\lpar 1 - \frac{1}{\rho_-} \rpar^{-1/2}\,
\lpar 1 - \frac{1}{\rho_+} \rpar^{-1/2}\,
\lpar \frac{\lambda}{4} \rpar^{1 + \mrs_1}
\nl
{}&\times&
\mrF^{(2)}_{\mrD}\lpar \frac{1}{2} - \mrs_2\,;\,\frac{1}{2}\,,\,\frac{1}{2}\,;\,\frac{9}{2} + \mrs_1\,;\,
\frac{1}{1 - \rho_-}\,,\,\frac{1}{1 - \rho_+} \rpar \spp
\eqa
Since $\rho_{\pm} > 1$ we can use the standard MB representation for $\mrF^{(2)}_{\mrD}$ and
no imaginary part will emerge.

In the region $1 < \lambda < 9$ we have $0 < \rho_0 < 1$ and $\rho_{\pm} \in \Cf$. We split the integration
in $[ 0 < \rho < \rho_0 ]$ ($\mrS_{<}$) and $[ \rho_0 < \rho < 1 ]$ ($\mrS_{>}$). The results are

\bqa
\mrS_{<}\lpar 1 < \lambda < 9 \rpar &=& - \int_{\mrL_1}\,\frac{\mrd \mrs_1}{\tip}\,
\frac{
\eG{ - \mrs_1}\,
\eGs{1 + \mrs_1}\,
\eG{1/2 + \mrs_1}\,
\eG{4 + \mrs_1}\,
}
{
\eG{3/2 + \mrs_1}\,
\eG{5 + 2*\mrs_1}
}
\nl
{}&\times&
\lpar \rho_0 \rpar^{4 + 2\,\mrs_1}\,
\lpar \frac{4}{\lambda} \rpar^{-1 - \mrs_1}\,
\mrF^{(2)}_{\mrD}\lpar
4 + \mrs_1\,;\,1 + \mrs_1\,,\,1 + \mrs_1\,;\,5 + 2\,\mrs_1\,;\,\frac{\rho_0}{\rho_-}\,,\,\frac{\rho_0}{\rho_+} \rpar \spp
\eqa

\bqa
\mrS_{>}\lpar 1 < \lambda < 9 \rpar &=& - 2\,\frac{\sqrt{\lambda}}{\pi}\,\int_{\mrL_1}\,\frac{\mrd \mrs_1}{\tip}\,
\frac{
\eG{ - \mrs_1}\,
\eG{1 + \mrs_1}\,
\eG{1/2 - \mrs_1}\,
\eGs{1/2 + \mrs_1}\,
}
{
\eG{3/2 + \mrs_1} 
}\,
\lpar \frac{\lambda}{4} \rpar^{\mrs_1}\,
\lpar \rho_0 \rpar^{3 + \mrs_1}\,
\lpar 1 - \rho_0 \rpar^{1/2}
\nl
{}&\times&
\mrF^{(3)}_{\mrD}\lpar
1 + \mrs_1\,;\,\frac{1}{2}\,,\,\frac{1}{2}\,,\, - 3 - \mrs_1\,;\,\frac{3}{2}\,;\,
\frac{\rho_0 - 1}{\rho_0 - \rho_-}\,,\,
\frac{\rho_0 - 1}{\rho_0 - \rho_+}\,,\,
\frac{\rho_0}{\rho_0 - 1} \rpar \spp
\eqa
\paragraph{Kite} \hspace{0pt} \\
In computing the kite diagram~\cite{Adams:2016xah,Remiddi:2016gno} we use the partial quadratization procedure described in 
\Bref{Passarino:2024ugq} and consider the following integral:
\bq
\mrI_{\mrK} = \int_0^1\,\mrd \mrx_1 \mrd \mrx_2\,\Bigl[ \mrQ(\mrx_1\,,\,\mrx_2) + \Delta \Bigr]^{-1} \spc \quad
\mrQ = \mra\,(\mrx_1 - \frac{1}{2})^2 + \mrb\,(\mrx_2 - \frac{1}{2})^2 +
\mrc\,(\mrx_1 - \frac{1}{2})\,(\mrx_2 - \frac{1}{2}) \spc
\eq
where the coefficients are
\bq
\mra = \rho_1^2\,\sigma_1 \spc \quad
\mrb = \rho_2^2\,\sigma_2 \spc \quad
\mrc = 2\,\rho_1\,\rho_2\,\rho_3 \spc
\eq
with $\rho_3 = 1 - \rho_1 - \rho_2$, $\sigma_\mrj = 1 - \rho_\mrj$; furthermore $0 \le \rho_1 \le 1$ and
$0 \le \rho_2 \le 1 - \rho_1$. The explicit expression for $\Delta$ depends on the internal masses; here
we discuss the case where $\mrm_3 = 0$ and $\mrm_\mrj = \mrm, \mrj \not= 3$. In this case we have
\bq
\Delta = \mrC\,\beta\,\sigma_3 \spc \qquad
\mrC = \lambda - \frac{1}{4} \spc \quad
\lambda = \frac{\mrm^2}{\mrs} \spc \quad
\beta = \rho_1\,\rho_3 + \rho_2\,\sigma_2 > 0 \spp
\eq
We want to study the kite integral below the normal threshold, \ie $\mrs < 4\,\mrm^2$. It follows that
$\mrC$ is positive and also $\mrQ > 0$ since the corresponding discriminant, $\mrc^2 - 4\,\mra\,\mrb$, is
negative. In this case we can use a MB splitting, \ie
\bq
\Bigl[ \mrQ(\mrx_1\,,\,\mrx_2) + \Delta \Bigr]^{-1} =
\int_{\mrL_1}\,\frac{\mrd \mrs_1}{\tip}\,\Delta^{- 1 - \mrs_1}\,\mrQ^{\mrs} \spp
\eq
After a change of variales, $\mrx^{\prime}_{1,2} = \mrx_{1,2} - 1/2$, we transform again into $[ 0\,,\, 1 ]^2$.
Using a sector decomposition we derive
\bq
\mrI_{\mrk} = \int_{\mrL_1}\,\frac{\mrd \mrs_1}{\tip}\,\mrJ_{\mrk} \spc \quad
\mrJ_{\mrK} = \frac{1}{2}\,4^{-\mrs_1}\,\Delta^{-1 - \mrs_1}\,\eB{-\mrs_1}{1 + \mrs_1}\,
\int_0^1\,\mrd \mrx_1 \mrd \mrx_2\,\lpar \mrJ_1 + \mrJ_2 \rpar \spc
\eq
\bqa
\mrJ_1 &=& \mrx_1^{1 + 2\,\mrs_1}\,\Bigl[
\lpar \mrb\,\mrx_2^2 + \mrc\,\mrx_2 + \mra \rpar^{\mrs_1} +
\lpar \mrb\,\mrx_2^2 - \mrc\,\mrx_2 + \mra \rpar^{\mrs_1} \Bigl] \spc
\nl
\mrJ_2 &=& \mrx_2^{1 + 2\,\mrs_1}\,\Bigl[
\lpar \mra\,\mrx_1^2 + \mrc\,\mrx_1 + \mrb \rpar^{\mrs_1} +
\lpar \mra\,\mrx_1^2 - \mrc\,\mrx_1 + \mrb \rpar^{\mrs_1} \Bigl] \spp
\eqa
Integration over $\mrx_1, \mrx_2$ gives
\bqa
\mrI_{\mrK} &=& \frac{1}{2}\,\int_{\mrL_1}\,\frac{\mrd \mrs_1}{\tip}\,\eG{ - \mrs_1}\,
\frac{\eG{1 + \mrs_1}\,\eG{2 + 2\,\mrs_1}}{\eG{3 + 2\,\mrs_1}}\,
4^{ - \mrs_1}\,\Delta^{-1 - \mrs_1}\,\lpar \mrH_1 + \mrH_2 \rpar \spc
\nl
\mrH_1 &=& \mra^{\mrs_1}\,\Bigl[
\mrF^{(2)}_{\mrD}\lpar 1\,;\, - \mrs_1\,,\, - \mrs_1\,;\,2\,;\,\mrr_5^{-1}\,,\,\mrr_6^{-1} \rpar +
\mrF^{(2)}_{\mrD}\lpar 1\,;\, - \mrs_1\,,\, - \mrs_1\,;\,2\,;\,\mrr_7^{-1}\,,\,\mrr_8^{-1} \rpar \spc
\nl
\mrH_2 &=& \mrb^{\mrs_1}\,\Bigl[
\mrF^{(2)}_{\mrD}\lpar 1\,;\, - \mrs_1\,,\, - \mrs_1\,;\,2\,;\,\mrr_1^{-1}\,,\,\mrr_2^{-1} \rpar +
\mrF^{(2)}_{\mrD}\lpar 1\,;\, - \mrs_1\,,\, - \mrs_1\,;\,2\,;\,\mrr_3^{-1}\,,\,\mrr_5^{-1} \rpar \spp
\eqa
The roots $\mrr_{\mrj}$ are given by
\bqa
\mrr_{1,2} &=& \frac{\rho_2}{\rho_1\,\sigma_1}\,\lpar \rho_3 \pm \mri\,\sqrt{\beta} \rpar \spc \qquad
\mrr_{3,4} =   \frac{\rho_2}{\rho_1\,\sigma_1}\,\lpar - \rho_3 \pm \mri\,\sqrt{\beta} \rpar \spc
\nl
\mrr_{5,6} &=& \frac{\rho_1}{\rho_2\,\sigma_2}\,\lpar \rho_3 \pm \mri\,\sqrt{\beta} \rpar \spc \qquad
\mrr_{7,8} =   \frac{\rho_1}{\rho_2\,\sigma_2}\,\lpar - \rho_3 \pm \mri\,\sqrt{\beta} \rpar \spp
\eqa
Since $\mrr_\mrj \in \Cf$ we can use the MB representation for the Lauricella functions. The complete kite integral
is:
\bq\mrK = \int_0^1 \mrd \rho_1\,\int_0^{1 - \rho_1}\,\mrd \rho_2\,\frac{\rho_1\,\rho_2}{\beta}\,\mrI_{\mrK} \spp
\eq
The only problem in performing the last two integrations is represented by the presence of 
$\sqrt{\beta}$. First
we change variable
\bq
\int_0^{1 - \rho_1}\,\mrd \rho_2\,\mrf(\rho_1\,\rho_2) = \sigma_1\,\int_0^1\,\mrd \rho_2\,
\mrf(\rho_1,\sigma_1\,\rho_2) \spp
\eq
Now we have $\sqrt{\omega^2 + \rho_2 - \rho_2^2}$ with $\omega^2 = \rho_1/\sigma_1$ and introduce a new 
variable $\mrt$ as follows:
\bq
\rho_2 = \frac{\mrt^2 - 2\,\omega\,\mrt}{\mrt^2 + 1} \spc \quad
\sqrt{\omega^2 + \rho_2 - \rho_2^2} = \frac{\omega\,\mrt^2 + \mrt - \omega}{\mrt^2 + 1} \spc \qquad
\int_0^1\,\mrd \rho_2 = 2\,\int_{2\,\omega}^{\infty}\,\mrd \mrt\,
\frac{\omega\,\mrt^2 + \mrt - \omega}{\mrt^2 + 1} \spp 
\eq
The integration over $\mrt$ requires the introduction of the following roots:
\bq
\mrt_{3\,\pm} = - 2\,\rho_1\,(1 \mp \mri\,\omega) \spc \qquad
\mrt_{4\,\pm} = 2\,(1 \pm \mri\,\omega) \spc
\eq
\bq
\mrt_{5\,-} = - 2\,\frac{\omega^2}{1 + \mri\,\omega} \spc \qquad \mrt_{5\,+} =  2\,\mri\,\omega \spc \qquad
\mrt_{6\,-} = - 2\,\frac{\omega^2}{1 - \mri\,\omega} \spc \qquad \mrt_{6\,+} = - 2\,\mri\,\omega \spp
\eq
We introduce four sets of $\mrb$ coefficients,
\[
\begin{array}{llll}
\mbox{Set 1} &&& \\
 \mrb_1 = 2 + \mrs_4 &
 \mrb_2 = 2 + \mrs_4 & 
 \mrb_3 = 1 + \mrs_1 & 
 \mrb_4 = 1 + \mrs_1 \\
 \mrb_5 = \mrs_2 &
 \mrb_6 = \mrs_2 &
 \mrb_7 = \mrs_3 &
 \mrb_8 = \mrs_3 \\
 \mrb_9 =  - \mrs_4 &
 \mrb_{10} =  - 2 - \mrs_4 - 2\,\mrs_3 - 2\,\mrs_2 &
 \mrb_{11} =  - 2 - \mrs_4 - 2\,\mrs_3 - 2\,\mrs_2 &  \\
\end{array}
\]
\[
\begin{array}{llll}
\mbox{Set 2} &&& \\
 \mrb_1 = 2 + \mrs_4 &
 \mrb_2 = 2 + \mrs_4 &
 \mrb_3 = 1 + \mrs_1 &
 \mrb_4 = 1 + \mrs_1 \\
 \mrb_5 = \mrs_3 &
 \mrb_6 = \mrs_3 &
 \mrb_7 = \mrs_2 &
 \mrb_8 = \mrs_2 \\
 \mrb_9 =  - \mrs_4 &
 \mrb_{10} =  - 2 - \mrs_4 - 2\,\mrs_3 - 2\,\mrs_2 &
 \mrb_{11} =  - 2 - \mrs_4 - 2\,\mrs_3 - 2\,\mrs_2 & \\
\end{array}
\]
\[
\begin{array}{llll}
\mbox{Set 3} &&& \\
 \mrb_1 = 2 + \mrs_4 - \mrs_3 - \mrs_2 + \mrs_1 &
 \mrb_2 = 2 + \mrs_4 - \mrs_3 - \mrs_2 + \mrs_1 &
 \mrb_3 = 1 + \mrs_1 &
 \mrb_4 = 1 + \mrs_1 \\
 \mrb_5 = \mrs_2 &
 \mrb_6 = \mrs_2 &
 \mrb_7 = \mrs_3 &
 \mrb_8 = \mrs_3 \\
 \mrb_9 =  - \mrs_4 &
 \mrb_{10} =  - 2 - \mrs_4 + \mrs_3 + \mrs_2 - 3\,\mrs_1 &
 \mrb_{11} =  - 2 - \mrs_4 + \mrs_3 + \mrs_2 - 3\,\mrs_1 & \\
\end{array}
\]
\[
\begin{array}{llll}
\mbox{Set 4} &&& \\
 \mrb_1 = 2 + \mrs_4 - \mrs_3 - \mrs_2 + \mrs_1 &
 \mrb_2 = 2 + \mrs_4 - \mrs_3 - \mrs_2 + \mrs_1 &
 \mrb_3 = 1 + \mrs_1 &
 \mrb_4 = 1 + \mrs_1 \\
 \mrb_5 = \mrs_3 &
 \mrb_6 = \mrs_3 &
 \mrb_7 = \mrs_2 &
 \mrb_8 = \mrs_2 \\
 \mrb_9 =  - \mrs_4 &
 \mrb_{10} =  - 2 - \mrs_4 + \mrs_3 + \mrs_2 - 3\,\mrs_1 &
 \mrb_{11} =  - 2 - \mrs_4 + \mrs_3 + \mrs_2 - 3\,\mrs_1 & \\
\end{array}
\]
There are six sets of $\mra, \mrc$ coefficients defined by
\bqa
\mrt^{\mra-1}\,(1 - \mrt)^{\mrc - \mra - 1} &=&
\mrt^{1+\mrs_4}\,(1 - \mrt)^{-1-\mrs_4-\mrs_3-\mrs_2+\mrs_1} \spc
\nl
\mrt^{\mra-1}\,(1 - \mrt)^{\mrc - \mra - 1} &=&
\mrt^{1+\mrs_4}\,(1 - \mrt)^{-1-\mrs_4+\mrs_3+\mrs_2-\mrs_1} \spc
\nl
\mrt^{\mra-1}\,(1 - \mrt)^{\mrc - \mra - 1} &=&
\mrt^{2+\mrs_4}\,(1 - \mrt)^{-1-\mrs_4-\mrs_3-\mrs_2+\mrs_1} \spc
\nl
\mrt^{\mra-1}\,(1 - \mrt)^{\mrc - \mra - 1} &=&
\mrt^{2+\mrs_4}\,(1 - \mrt)^{-1-\mrs_4+\mrs_3+\mrs_2-\mrs_1} \spc
\nl
\mrt^{\mra-1}\,(1 - \mrt)^{\mrc - \mra - 1} &=&
\mrt^{\mrs_4}\,(1 - \mrt)^{-1-\mrs_4-\mrs_3-\mrs_2+\mrs_1} \spc
\nl
\mrt^{\mra-1}\,(1 - \mrt)^{\mrc - \mra - 1} &=&
\mrt^{\mrs_4}\,(1 - \mrt)^{-1-\mrs_4+\mrs_3+\mrs_2+\mrs_1} \spp
\eqa
Therefore the integrals to be computed are,
\bqa
\mrK_{\mrj\,,\,\mrk} &=&
\int_0^1\,\mrd \mrt\,\mrt^{\mra_{\mrj} - 1}\,(1 - \mrt)^{\mrc_{\mrj} - \mra_{\mrj} - 1}\,
   \mrB_{\mrk} \spc
\nl
\mrB_{\mrk} &=&
   (1 - \mrt\,\mrt_{3\,-}^{-1})^{ - \mrb_{1,\mrk}}\,
   (1 - \mrt\,\mrt_{3\,+}^{-1})^{ - \mrb_{2,\mrk}}\,
   (1 - \mrt\,\mrt_{4\,-}^{-1})^{ - \mrb_{3,\mrk}}
\nl
{}&\times&
   (1 - \mrt\,\mrt_{4\,+}^{-1})^{ - \mrb_{4,\mrk}}\,
   (1 - \mrt\,\mrt_{5\,-}^{-1})^{ - \mrb_{5,\mrk}}\,
   (1 - \mrt\,\mrt_{5\,+}^{-1})^{ - \mrb_{6,\mrk}}\,
   (1 - \mrt\,\mrt_{6\,-}^{-1})^{ - \mrb_{7,\mrk}}
\nl
{}&\times&
   (1 - \mrt\,\mrt_{6\,+}^{-1})^{ - \mrb_{8,\mrk}}\,
   (1 + \frac{1}{4}\,\mrt\,\omega^{-2})^{ - \mrb_9}\,
   (1 - \frac{1}{4}\,\mri\,\mrt\,\omega^{-1})^{ - \mrb_{10,\mrk}}\,
   (1 + \frac{1}{4}\,\mri\,\mrt\,\omega^{-1})^{ - \mrb_{11,\mrk}} \spc
\eqa
The result is
\bqa
\mrK &=& \Bigl[ \prod_{\mrj=1}^{4}\,\int_{\mrL_{\mrj}}\,\frac{\mrd \mrs_\mrj}{\tip} \Bigr]\,
\Gamma_{1{-}4}\,\int_0^1 \mrd \rho_1\,\Bigl[
\mrQ_4\,(\mrK_{1,1} + \mrK_{1,3} )\,
+ \mrQ_3\,(\mrK_{2,2} + \mrK_{2,4} )
\nl
{}&-& \frac{1}{2}\,\mrQ_4\,(\mrK_{3,1} + \mrK_{3,3})\,
- \frac{1}{2}\,\mrQ_3\,(\mrK_{4,2} + \mrK_{4,4})\,
+ \frac{1}{2}\,\mrQ_1\,(\mrK_{5,1} + \mrK_{5,3})\,
+ \frac{1}{2}\,\mrQ_2\,(\mrK_{6,2} + \mrK_{6,4}) \Bigr] \spp
\eqa
Auxiliary quantities are $\Gamma_{1{-}4}= \mrN/\mrD$,
\bqa
\mrN &=&
\eG{ - \mrs_2}\,
\eG{ - \mrs_3}\,
\eG{ - \mrs_4}\,
\eG{1 + \mrs_1}\,
\eG{2 + 2\,\mrs_1}\,
\eG{\mrs_2 - \mrs_1}\,
\eG{\mrs_3 - \mrs_1}\,
\eG{1 + \mrs_3 + \mrs_2}\,
\eG{2 + \mrs_4 + \mrs_1} \spc
\nl
\mrD &=&
\eG{ - \mrs_1}\,
\eG{2 + \mrs_1}\,
\eG{3 + 2\,\mrs_1}\,
\eG{2 + \mrs_3 + \mrs_2} \spc
\eqa
\bqa
\mrQ_1 &=& 4^{-\mrs_1}\,\mrC^{-1-\mrs_1}\,( - \mri\,\omega)^{-\mrs_2}\,(\mri\,\omega)^{-\mrs_3}\,
\rho_1^{-1+\mrs_3+\mrs_2}\,(1 - \rho_1)^{-\mrs_3-\mrs_2+\mrs_1} \spc
\nl
\mrQ_2 &=& 4^{-\mrs_1}\,\mrC^{-1-\mrs_1}\,( - \mri\,\omega)^{-\mrs_2}\,(\mri\,\omega)^{-\mrs_3}\,
\rho_1^{-1+\mrs_1} \spc
\nl
\mrQ_3 &=& 4^{-1-\mrs_1}\,\mrC^{-1-\mrs_1}\,\,( - \mri\,\omega)^{-\mrs_2}\,(\mri\,\omega)^{-\mrs_3}\,
\omega^{-2}\,\rho_1^{-1+\mrs_1} \spc
\nl
\mrQ_4 &=& 4^{-1-\mrs_1}\,\mrC^{-1-\mrs_1}\,\,( - \mri\,\omega)^{-\mrs_2}\,(\mri\,\omega)^{-\mrs_3}\,
\omega^{-2}\,\rho_1^{-1+\mrs_3+\mrs_2}\,(1 - \rho_1)^{-\mrs_3-\mrs_2+\mrs_1} \spp
\eqa
The integrals give rise to $\mrF^{(11)}_{\mrD}\lpar \mra_\mrj\,;\,\mathbf{b}_{\mrk}\,;\,\mrc_{\mrj}\,;\,\mathbf{z} \rpar$ 
Lauricella functions,
We can use a MB representation since all conditions are satisfied. For the remaining $\rho_1$ integration we
transform as follows:
\bq
\int_0^1\, \mrd \rho_1\,\mrf(\rho_1) = 2\,\int_0^{\infty}\,\mrd \omega\,\frac{\omega}{(\omega^2 + 1)^2}\,
\mrf(\omega) \spc \qquad \rho_1 = \frac{\omega^2}{\omega^2 + 1} \spc
\eq
followed by
\bq
\int_0^{\infty}\, \mrd \omega = \int_0^1\,\frac{\mrd \mru}{(\mru - 1)^2} \spc \qquad
\omega = \frac{\mru}{1 - \mru} \spc
\eq
obtaining another set of Lauricella functions that can be transformed into Fox functions.
\subsection{Real corrections \label{rea}}
We present two examples related to radiative processes.
\paragraph{Real emission: $1$} \hspace{0pt} \\
Fox functions can also be used to compute real emission; for the sake of simplicity we consider the process
\bq
\mrH(\mrP) \to \PGg(\mrq_1) + \PAQb(\mrq_3) + \PQb(\mrq_2) \spc
\eq
at lowest order. The fully massive $\mrd\,${-}dimensional phase space can be written as follows:
\bq
\int\,\mrd\mrP\mrS = \upphi(\mrd)\,\int_0^{\beta^2}\,\mrd \mry\,\int_{\mrx_-}^{\mrx_+}\,\mrd \mrx\,
\mrQ^{\mrd/2 - 2}(\mrx\,,\,\mry) \spc \quad
\mrQ= (1 - \mry)\,\Bigl[ \mrx\,\mry - (\mrx - \mrr)^2 \Bigr] - 4\,\mrr\,\mry \spc
\eq
where we have introduced
\bq
\mrM^2_{\mrH}\,\mrx = - (\mrq_2 + \mrq_1)^2 \spc \quad
\mrM^2_{\mrH}\,(1 + \mry) = - (\mrP - \mrq_1)^2 \spc
\eq
\bq
\mrr = \frac{\mrm^2_{\PQb}}{\mrM^2_{\mrH}} \spc \quad
\beta^2 = 1 - 4\,\mrr \spc \quad
\mrx_{\pm} = \mrr + \frac{1}{2}\,\mry\,\Bigl[ 1 \pm \mrR(\mry) \Bigr] \spc \quad
\mrR^2(y)= 1 - 4\,\frac{\mrr}{1 - \mry} \spp
\eq
and an overall factor
\bq
\upphi(\mrd) = 2^{5 - \mrd}\,\pi^{\mrd - 2}\,\frac{\mrM_{\mrH}^{2\,\mrd - 10}}{\eG{\mrd - 2}} \spp
\eq
Next we compute the amplitude and obtain
\bq
\sum_{\spin}\,\mid \mrA_{\myLO} \mid^2 = \frac{1}{9}\,\mrg^4\,\sin^2(\theta)\,\frac{\mrm^2_{\PQb}}{\MHs}\,
\mcA \spc
\eq
where $\mrg$ is the $SU(2)$ coupling constant and $\theta$ is the weak{-}mixing angle. Furthermore
\bqa
\mcA &=&
       - 2\,\frac{\mrr}{(\mrx - \mrr)^2}\,(1 - 4\,\mrr)
       + 2\,\frac{1}{\mrx - \mrr}\,\frac{1}{\mry}\,(1 - 6\,\mrr + 8\,\mrr^2) 
       - 2\,\frac{1}{\mrx - \mrr}\,(1 - 4\,\mrr) 
       + \frac{1}{2}\,\frac{1}{\mrx - \mrr}\,y\,(2 + \ep) 
\nl
{}&-& 2\,\frac{\mrr}{(\mrx - \mry - \mrr)^2}\,(1 - 4\,\mrr)
       - 2\,\frac{1}{\mrx - \mry - \mrr}\,\frac{1}{\mry}\,(1 - 6\,\mrr + 8\,\mrr^2) 
       + 2\,\frac{1}{\mrx - \mry - \mrr}\,(1 - 4\,\mrr) 
       - \frac{1}{2}\,\frac{1}{\mrx - \mry - \mrr}\,y\,(2 + \ep) \spp 
\label{Amphbbg}
\eqa
The computation proceeds through the following steps:
\begin{enumerate}

\item $\mrx^{\prime} = \mrx - \mrr$, giving
$\mrx^{\prime}_{\pm} = \frac{1}{2}\,(1 \pm \mrR )$,

\item $\mrx^{\prime} = \mry\,\mrx^{\prime\prime}$, giving 
$\mrx^{\prime\prime}_{\pm} = \mrX_{\pm} = \frac{1}{2}\,(1 \pm \mrR)$ \spp

\end{enumerate}
In this way we obtain $\mrQ = (1 - \mry)\,(\mrx - \mrX_-)\,(\mrX_+ - \mrx)$, and change again variables
$\mrx = \mrX_+ - \mrx^{\prime}$ so that
\bq
\int_{\mrX_-}^{\mrX_+}\,\mrd \mrx \quad \to \quad ( \mrX_+ - \mrX_- )\,\int_0^1\,\mrd \mrx \spp
\eq
At this point we have to compute the following integral:
\bq
  \mrI_{\pm} = \int_0^1 \mrd \mrx\,
  \mrx^{\mra_1}\,(1 - \mrx)^{\mra_2}\,\Bigl[
  1 - \frac{\mrX_+ - \mrX_-}{\mrX_{\pm}}\,\mrx \Bigr]^{\mra_3} \spc
  \eq
with the following result:
\bq
  \mrI_{\pm} = \eB{\mra_1 + 1}{\mra_2 + 1}  \,
  \ghyp{2}{1}\Bigl( - \mra_3\,;\,\mra_1 + 1\,,\,\mra_1 + \mra_2 + 2\,;\,
  \frac{\mrX_+ - \mrX_-}{\mrX_{\pm}}\Bigr)
\eq
Since $\mrX_+ - \mrx_- = \mrR$ we have the following argument for the
hypergeometric function:
\bq
 \zeta_{+} = 2\,\frac{\mrR}{1 + \mrR} \spc \qquad
0 < \zeta_{+} < 1 \spc
\eq
\bq
    \zeta_{-} = 2\,\frac{\mrR}{1 - \mrR} \spc \quad
    \zeta_{-} > 1 \quad \hbox{for} \quad
    \mrm_{\PQb} << \mrM_{\mrH} \spp
\eq
Therefore we transform the hypergeometric funcion of argument $\zeta_{+}$
according to
\bq
   \ghyp{2}{1}(\mra\,\,\mrb\,;\,\mrc\,;\,\zeta_1) =
   (1 - \zeta_1)^{ - \mra}\,
   \ghyp{2}{1}(\mra\,\,\mrc - \mra\,;\,\mrc\,;\,
   \frac{\zeta_1}{\zeta_1 - 1}) \spc
\eq
\bq
    1 - \zeta_{+} = \frac{1 - \mrR}{1 + \mrR} \spc \quad
\frac{\zeta_{+}}{\zeta_{+} - 1} = 2\,\frac{\mrR}{\mrR -1}
\eq
Since $0 < y < 1 - 4\,\mrr$ we have $0 < \mrR < \beta$,
\ie  $0 < \mrR < 1$.At this point we can use the MB representation for the
hypergeometric function . The presence of $\mrR(\mry)$ requires another
change of variables:
\bq
   \mry =1 - 4\,\frac{\mrr}{1 -\mrz^2} \spc \quad\hbox{corresponding to}
\quad
   \int_0^{\beta^2}\, \mrd \mry = 8\,\mrr\,
   \int_0^{\beta}\,\mrd
    \mrz\,\frac{\mrz}{1 - \mrz^2} \spp
\eq
The final change of variables will be $\mrz = \beta\,\mrz^{\prime}$ which
allows us to obtain a result which is a combination of $\mrF^{(3)}_\mrD$
Lauricella functions of argument $ - 1\,,\, - \beta\,,\, \beta$.
It is convenient to transform the Lauricella function according to
\bqa
  \mrF^{(3)}_{\mrD} \Bigl(
  \mra\,;\,\mathbf{b}\,;\,\mrc\,;\,- 1\,,\, - \beta\,,\,\beta\Bigr) &=&
  2^{\mrc - \mra - \mrb_1}\,(1 - \beta)^{ - \mrb_2}\,
  (1 - \beta)^{ - \mrb_3}
\nl
{}&\times&
  \mrF^{(3)}_{\mrD}\Bigl(
  \mrc - \mra\,;\,\mrc - \sum_\mrj\,
\mrb_\mrj\,,\,\mrb_2\,,\,\mrb_3\,;\,\mrc\,;\, - 1\,,\, 
- \frac{1 -\beta}{1 + \beta}\,,\, - \frac{1 + \beta}{1 - \beta} \Bigr) \spp
\eqa
This equation should be interpreted with due care since
$\mrc - \sum_\mrj\,\mrb_\mrj$ could be negative.
Since $\beta < 1$ all arguments of the transformed Lauricella function are negative and we can introduce
the corresponding MB representation, performing the last integration.
In order to present the final result we need to introduce auxiliary quantities:
\bqa
\Gamma_{-}(\mrs_1,\mrs_2,\mrs_3)  &=& \eG{ - \mrs_1}\,\eG{ - \mrs_2}\,\eG{ - \mrs_3} \spc
\nl
\Gamma_1(\mrn_1,\mrn_2,\mrn_3,\mrn_4,\mrn_5,\mrn_6) &=& 
\eG{\mrn_1 + \frac{1}{2}\,\ep + \mrs_1}
         \,\eG{\mrn_2 + \frac{3}{2}\,\ep + \mrs_2}
         \,\eG{\mrn_3 + \mrs_1}
\nl {}&\times&
         \,\eG{\mrn_4 + \ep + \mrs_3 + \mrs_2}
         \,\eG{\mrn_5 + \ep + \mrs_1}
         \,\eG{\mrn_6 + \frac{3}{2}\,\ep + \mrs_3 + \mrs_1}
         \,\Gamma_{-}(\mrs_1,\mrs_2,\mrs_3) \spc
\nl    
\Gamma_2(\mrn_1,\mrn_2,\mrn_3,\mrn_4,\mrn_5) &=& 
\eG{\mrn_1 + \frac{1}{2}\,\ep + \mrs_1}
         \,\eG{\mrn_2 + \frac{3}{2}\,\ep + \mrs_2}
         \,\eG{\mrn_3 + \mrs_1}
\nl {}&\times&
         \,\eG{\mrn_4 + \ep + \mrs_1}
         \,\eG{\mrn_5 + \frac{3}{2}\,\ep + \mrs_3 + \mrs_1}
         \,\eG{\ep + \mrs_3 + \mrs_2}
         \,\Gamma_{-}(\mrs_1,\mrs_2,\mrs_3)  \spc
\nl
\Gamma_3(\mrn_1,\mrn_2,\mrn_3,\mrn_4,\mrn_5) &=&
\eG{\mrn_1 + \frac{1}{2}\,\ep + \mrs_1}
         \,\eG{\mrn_2 + \frac{3}{2}\,\ep + \mrs_2 + \mrs_1}
         \,\eG{\mrn_3 + \mrs_1}
\nl {}&\times&
         \,\eG{\mrn_4 + \ep + \mrs_3 + \mrs_2}
         \,\eG{\mrn_5 + \frac{3}{2}\,\ep + \mrs_3 - \mrs_1}
         \,\Gamma_{-}(\mrs_1,\mrs_2,\mrs_3)  \spc
\nl
\Gamma_4(\mrn_1,\mrn_2,\mrn_3,\mrn_4) &=&
\eG{\mrn_1 + \frac{1}{2}\,\ep + \mrs_1}
         \,\eG{\mrn_2 + \frac{3}{2}\,\ep + \mrs_2 + \mrs_1}
         \,\eG{\mrn_3 + \mrs_1}
\nl {}&\times&
         \,\eG{\mrn_4 + \frac{3}{2}\,\ep + \mrs_3 - \mrs_1}
         \,\eG{\ep + \mrs_3 + \mrs_2}
         \,\Gamma_{-}(\mrs_1,\mrs_2,\mrs_3) \spc
\nl
%
%
\Gamma^{-1}_5(\mrn_1,\mrn_2,\mrn_3,\mrn_4) &=&
\eG{\mrn_1 + \frac{3}{2}\,\ep + \mrs_1}
         \,\eG{\mrn_2 + \ep + \mrs_1}
         \,\eG{\mrn_3 + \frac{3}{2}\,\ep, - \mrs_1}
         \,\eG{\mrn_4 + 2\,\ep + \mrs_3 + \mrs_2} \spc 
\nl
\Gamma^{-1}_6(\mrn_1,\mrn_2,\mrn_3) &=&
\eG{\mrn_1 + \ep + \mrs_1}
         \,\eG{\mrn_2 + \frac{3}{2}\,\ep + \mrs_1}
         \,\eG{\mrn_3 + 2\,\ep + \mrs_3 + \mrs_2 + \mrs_1}  
\eqa
\bqa
\mrQ_1(\mrs_1,\mrs_2,\mrs_3,\beta) &=&
\mrP(\mrs_1,2\, - \frac{\beta}{1 - \beta}\,(1 + \delta))\,
\mrP(\mrs_2,\frac{1 - \beta}{1 + \beta})\,
\mrP(\mrs_3,\frac{1 + \beta}{1- \beta}) \spc
\nl
\mrQ_2(\mrs_1,\mrs_2,\mrs_3,\beta) &=&
\mrP(\mrs_1,\frac{1 - \beta}{1 + \beta})\,
\mrP(\mrs_2,\frac{1 - \beta}{1 + \beta})\,
\mrP(\mrs_3,\frac{1 + beta}{1 - \beta}) \spc
\eqa
\bqa
\mrP_1(\mrn_1,\mrn_2,\mrn_3,\beta) &=&
\mrP(\mrn_1 + \frac{1}{2}\,\ep,1 - \beta)
         \,\mrP(\mrn_2 + \frac{1}{2}\,\ep,1 + \beta)
         \,\mrP(\mrn_3 + 3\,\ep,\beta)
         \,\mrP( - \frac{3}{2}\,\ep,1 - \beta\,1 + \beta) \spc 
\nl
\mrP_2(\mrn_1,\mrn_2,\beta) &=&
\mrP(\mrn_1 + \frac{1}{2}\,\ep,1 - \beta)
         \,\mrP(\mrn_2 + 3\,\ep,\beta)
         \,\mrP(\frac{1}{2}\,\ep,1 + \beta)
         \,\mrP( - \frac{3}{2}\,\ep,1 - \beta\,1 + \beta) \spc
\nl
\mrP_3(\mrn_1,\mrn_2,\beta) &=&
\mrP(\mrn_1 + \frac{1}{2}\,\ep,1 + \beta)
         \,\mrP(\mrn_2 + 3\,\ep,\beta)
         \,\mrP(\frac{1}{2}\,\ep,1 - \beta)
         \,\mrP( - \frac{3}{2}\,\ep,1 - \beta\,1 + \beta) \spc
\eqa
where $\mrP(\mrc\,,\,\mrz)= \mrz^{\mrc}$.
The result is
\bq
\int\,\mrd\mrP\mrS\,\sum_{\spin}\,\mid \mrA_{\myLO} \mid^2 =
\frac{1}{9}\,\mrg^4\,\sin^2(\theta)\,\frac{\mrm^2_{\PQb}}{\MHs}\,4^{3/2\,\ep}\,\phi(4 + \ep)\,
\Bigl[ \prod_{\mrj=1}^{3}\,\int_{\mrL_\mrj}\,\frac{\mrd \mrs_\mrj}{2\,\mri\,\pi} \Bigr]\,
\Bigl[ \mrQ_1\,\mrJ_1 + \mrQ_2\,\mrJ_2 \Bigr] \spp
\eq
\bqa
\mrJ_1 &=&
        2\,\frac{\eG{1 + \ep/2}}{\eG{1 + 3/2\,\ep}} \, 
          \Bigl\{
           \Bigl[ \Gamma_1(1,1,1,1,3,2)\,\Gamma_6(2,2,4) 
                 + \Gamma_2(1,1,1,3,2)\,\Gamma_6(2,2,3) 
            \Bigr]
\nl
{}&\times&
               \Bigl[\mrP_1(1,2,1,\beta) + 2\,\mrP_2(-1,1,\beta) - 3\,\mrP_3(1,1,\beta) \Bigr]
\nl
{}&-&           \Bigl[2\,\Gamma_1(1,1,2,1,3,3)\,\Gamma_6(2,3,4) 
               + \Gamma_1(1,1,2,2,3,3)\,\Gamma_6(2,3,5) 
               + \Gamma_2(1,1,2,3,3)\,\Gamma_6(2,3,3) 
            \Bigr]\,\mrP_1(-1,1,2,\beta)
          \Bigr\}
\nl
{}&-& 16\,\frac{\eG{1 + \ep/2}}{\eG{2 + 3/2\,\ep}} \,
          \Bigl[ \Gamma_1(1,2,1,1,3,3)\,\Gamma_6(2,3,4) 
           + \Gamma_1(1,2,1,2,3,3)\,\Gamma_6(2,3,5) \Bigr]\,
           \Bigl[ \mrP_1(-2,-1,3,\beta) - \mrP_2(-1,3,\beta) \Bigr]
\nl          
{}&+& \frac{\eG{1 + \ep/2}}{\eG{3 + 3/2\,\ep}} \, 
          \Bigl\{
           4\,\Bigl[ \Gamma_1(1,3,1,1,3,4)\,\Gamma_6(2,4,4) 
               - \Gamma_1(1,3,1,1,5,4)\,\Gamma_6(2,4,6) 
\nl
{}&+&              3\,\Gamma_1(1,3,1,2,3,4)\,\Gamma_6(2,4,5) 
             -   \Gamma_1(1,3,1,2,5,4)\,\Gamma_6(2,4,7) 
\nl
{}&+&
              3\,\Gamma_1(1,3,1,3,3,4)\,\Gamma_6(2,4,6) 
              + \Gamma_1(1,3,1,4,3,4)\,\Gamma_6(2,4,7) \Bigr]\,\mrP_1(-3,-2,5,\beta)\,(2 + \ep)
          \Bigr\} \spc
\eqa
\bqa
\mrJ_2 &=&
      \eG{1 + \ep/2}\,\eG{3 + \ep}\,
          \Bigl\{
           2\,\Bigl[ \Gamma_3(1,1,1,1,2)\,\Gamma_5(1,2,2,4) 
               + \Gamma_4(1,1,1,2)\,\Gamma_5(1,2,2,3) \Bigr]\,
\nl
{}&\times&
              \Bigl[\mrP_1(1,2,1,\beta) + 2\,\mrP_2(-1,1,\beta) - 3\,\mrP_3(1,1,\beta) \Bigr]
\nl
{}&-&
           2\,\Bigl[ 2\,\Gamma_3(1,1,2,1,3)\,\Gamma_5(1,2,3,4) 
                 + \Gamma_3(1,1,2,2,3)\,\Gamma_5(1,2,3,5)
                 + \Gamma_4(1,1,2,3)\,\Gamma_5(1,2,3,3) \Bigr]\,
              \mrP_1(-1,1,2,\beta)
\nl
{}&-&
           16\,\Bigl[ \Gamma_3(1,2,1,1,3)\,\Gamma_5(2,2,3,4) 
                + \Gamma_3(1,2,1,2,3)\,\Gamma_5(2,2,3,5) \Bigr]\,
              \mrP_1(-2,-1,4,\beta)
\nl
{}&+&
           4\,\Bigl[ \Gamma_3(1,3,1,1,4)\,\Gamma_5(3,2,4,4) 
             + 3\,\Gamma_3(1,3,1,2,4)\,\Gamma_5(3,2,4,5)
\nl
{}&+&
              3\,\Gamma_3(1,3,1,3,4)\,\Gamma_5(3,2,4,6) 
             + \Gamma_3(1,3,1,4,4)\,\Gamma_5(3,2,4,7) \Bigr]\,
              \mrP_1(-3,-2,5,\beta)\,(2 + \ep)
          \Bigr\}
\nl
{}&-& 4\,\eG{1 + \ep/2}\,\eG{5 + \ep}\, 
          \Bigl[ \Gamma_3(1,3,1,1,4)\,\Gamma_5(3,2,4,6) 
               + \Gamma_3(1,3,1,2,4)\,\Gamma_5(3,2,4,7) \Bigr]\,
\nl
{}&\times&
         \mrP_1(-3,-2,5,\beta)\,(2 + \ep) \spp
\eqa          
For infrared divergent integrals the role of $\ep$ is crucial and the
limit $\ep \to 0$ should be taken with due care. As an example, consider
\bq
 \mrJ = \int_0^1 \mrd \mrx\,\mrx^{\ep - 1}\,(1 - \mrx)^{\mrc - \ep -
1}\,(1 - \mrz\,\mrx)^{ - \mra} \spp
\eq
The integral diverges for $\ep \to 0$. We can write
\bq
  \mrI = \Gamma(\ep)\,\frac{\Gamma(\mrc -
\ep)}{\Gamma(\mrc)}\,\ghyp{2}{1}(\mra\,,\,\ep\,;\,\mrc\,;\,\mrz) \spp
\eq
Let us now introduce the MB representation of the result:
\bq
   \mrI =  \frac{\Gamma(\mrc - \ep)}{\Gamma(\mra)}\,
   \intMB\,\Gamma( - \mrs)\,\frac{\Gamma(\ep + \mrs)\,\Gamma(\mra +
\mrs)}{\Gamma(\mrc + \mrs)}\,( - \mrz)^{\mrs} \spp
\eq
If we set $\ep = 0$ we will have a divergence hidden in the MB integral;
indeed for $\ep = 0$ we cannot separate the poles, even distorting the
contour. Therefore, it is more convenient to use a contiguity relation
\bq
(\mrc - \mrb - 1)\,\ghyp{2}{1} = - \Bigl[ \mra\,\mrz - \mrc + (\mrb + 1)\,(2 -
\mrz) \Bigr]\,\ghyp{2}{1}(\mrb + 1) - (\mrb + 1)\,(\mrz - 1)\,\ghyp{2}{1}(\mrb + 2) \spp
\eq
Next we introduce the corresponding MB representations obtaining $\mrI =
\mrI_1 + \mrI_2$
\bqa
\mrI_1 &=& \mrH_2\,(1 + \ep)\,\intMB\,\Gamma(-\mrs)\,\frac{\Gamma(\mra
+ \mrs)\,\Gamma(2 + \ep + \mrs)}{\Gamma(\mrc + \mrs)}\,( - \mrz)^{\mrs}
\nl
{}&-& \mrH_1\,(2 - \mrc + 2\,\mrs)\,
     \intMB\,\Gamma( - \mrs)\,\frac{\Gamma(\mra + \mrs)\,\Gamma(1 + \ep +
\mrs)}{\Gamma(\mrc + \mrs)}\,( - \mrz)^{\mrs} \spp
\eqa
\bqa
\mrI_2  &=& \mrH_2\,(1 + \ep)\,\intMB\,\Gamma(-\mrs)\,\frac{\Gamma(\mra
+ \mrs)\,\Gamma(2 + \ep + \mrs)}{\Gamma(\mrc + \mrs)}\,( - \mrz)^{\mrs +
1} 
\nl
{}&-& \mrH_1\,(1 - \mra + \mrs)\,
     \intMB\,\Gamma( - \mrs)\,\frac{\Gamma(\mra + \mrs)\,\Gamma(1 + \ep +
\mrs)}{\Gamma(\mrc + \mrs)}\,( - \mrz)^{\mrs + 1} \spp
\eqa
\bq
\mrH_1 = \frac{\Gamma(\ep)\,\Gamma(\mrc - \ep)}{\Gamma(1 +
\ep)\,\Gamma(\mra)}\,\frac{1}{\mrc - 1 - \ep} \spc \qquad
\mrH_2 = \frac{\Gamma(\ep)\,\Gamma(\mrc - \ep)}{\Gamma(2 +
\ep)\,\Gamma(\mra)}\,\frac{1}{\mrc - 1 - \ep} \spp
\eq
The procedure can be generalized, for instance using
\bqa
{}&{}& \mrF^{(2)}_{\mrD}(\mra\,;\,\mrb_1\,,\,\mrb_2\,;\,\mrc\,;\,\mrz_1\,,\mrz_2) =
       \mrF^{(2)}_{\mrD}(\mra+1\,;\,\mrb_1\,,\,\mrb_2\,;\,\mrc\,;\,\mrz_1\,,\mrz_2)
\nl 
{}&-& \frac{\mrb_1}{\mrc}\,\mrF^{(2)}_{\mrD}(\mra+1\,;\,\mrb_1+1\,,\,\mrb_2\,;\,\mrc+1\,;\,\mrz_1\,,\mrz_2) -
      \frac{\mrb_2}{\mrc}\,\mrF^{(2)}_{\mrD}(\mra+1\,;\,\mrb_1\,,\,\mrb_2+1\,;\,\mrc+1\,;\,\mrz_1\,,\mrz_2) \spp
\eqa
Let us consider only one term in \eqn{Amphbbg}, \ie
\bq
\mcA^0 = 2\,\frac{1}{\mrx - \mrr}\,\frac{1}{\mry}\,(1 - 6\,\mrr + 8\,\mrr^2) \spp
\eq
The infrared divergent part of
\bq
\int\,\mrd\mrP\mrS\,\sum_{\spin}\,\mid \mrA^0_{\myLO} \mid^2 =
\frac{1}{9}\,\mrg^4\,\sin^2(\theta)\,\frac{\mrm^2_{\PQb}}{\MHs}\,4^{3/2\,\ep}\,\phi(4 + \ep)\,\mcA^0 \spc
\eq
is given by
\bqa
\Bigl[ \int\,\mrd\mrP\mrS\,\sum_{\spin}\,\mid \mrA^0_{\myLO} \mid^2 \Bigr]_{\mdiv} &=&
\Bigl[ \frac{1}{\ep} + \uppsi(1) \Bigr]\,\beta^3\,\frac{1 + \beta^2}{1 - \beta}\,
\int_{\mrL}\,\frac{\mrd \mrs}{2\,\mri\,\pi}\,\mrz^{\mrs}\,\eG{ - \mrs}\,\eG{1 + \mrs}\,
\nl
{}&\times& \Bigl\{
 - \frac{2}{3}\,\frac{\eG{1 + \mrs}\,\eG{1 + \mrs}\,\eG{3 - \mrs}}{\eG{2 - \mrs}\,\eG{2 + \mrs}}\,
\mrF^{(2)}_{\mrD}(1\,;\,1 + \mrs\,,\,3 - \mrs\,;\,4\,;\, - \mrz\,,\, - \mrz^{-1})
\nl
{}&+& 
  2\,\frac{\eG{1 + \mrs}\,\eG{1 + \mrs}}{\eG{2 + \mrs}}\,
\mrF^{(2)}_{\mrD}(1\,;\,1 + \mrs\,,\,2 - \mrs\,;\,3\,;\, - \mrz\,,\, - \mrz^{-1})
\nl
{}&-&
 \frac{2}{3}\,\eG{1 + \mrs}\,\eG{1 + \mrs}\,
\mrF^{(2)}_{\mrD}(1\,;\,2 + \mrs\,,\,2 - \mrs\,;\,4\,;\, - \mrz\,,\, - \mrz^{-1}) \} \spc
\eqa
where $\mrz = (1 - \beta)/(1 + \beta)$.
\paragraph{Real emission: $2$} \hspace{0pt} \\
The next example refers to the process 
\bq
\mrH(\mrP) \to \Pep(\mrq_2) + \Pem(\mrq_1) + \PGg(\mrk) \spc
\eq
Invariants describing the process are
\bq
\mrs = - (\mrP - \mrk)^2 = - (\mrq_1 + \mrq_2)^2 \spc \qquad
\mru = - (\mrP - \mrq_1)^2 = - (\mrq_2 + \mrk)^2 \spc
\eq
with a phase space described by
\bq
4\,\mrm^2 \le \mrs \le \MHs \spc \qquad \mru_{-} \le \mru \le \mru_{+} \spc \quad
\mru_{\pm} = \MHs + \mrm^2 - \frac{1}{2}\,(\MHs + \mrs) \pm \frac{1}{2}\,(\MHs - \mrs)\,\lambda \spc
\eq
where $\lambda^2 = 1 - 4\,\mrm^2/\mrs$.
We limit our analysis to the interference beween the LO amplitude and one NLO diagram, the box with propagators
\bq
\mrq^2 + \mrm^2 \spc \quad
(\mrq - \mrk)^2 + \mrm^2 \spc \quad
(\mrq - \mrq_1 - \mrk)^2 \spc \quad
(\mrq - \mrq_1 - \mrq_2 - \mrk)^2 + \mrm^2 \spp
\eq
The interference can be written in terms of three{-}point and four{-}point scalar functions. It will be
enough to consider the following term:
\bqa
\mrD &=& \eG{2 - \frac{1}{2}\,\ep}\,\pi^{\ep/2}\,  
\int_0^1 \mrd \mrx_1\,\int_0^{\mrx_1} \mrd \mrx_2 \,\int_0^{\mrx_2}\,
\mrV^{ - 2 + \ep/2}(\mrx_1\,,\,\mrx_2\,,\,\mrx_3)\, \mrd \mrx_3 \spc
\nl
\mrV &=&  
\mrm^2\,(\mrx_2^2 + \mrx_3^2) +
(\mrt - \mrm^2)\,\mrx_1\,\mrx_2 +
(\mru - \mrm^2)\,\mrx_1\,\mrx_3 +
(\mrs - 2\,\mrm^2)\,\mrx_2\,\mrx_3 -
(\mrt + \mrm^2)\,\mrx_2 
\nl
{}&+&
(\mrt + \mrm^2 + \MHs)\,\mrx_3 +
\mrm^2 \spp
\eqa
We can perform the $\mrx_1$ integration giving two terms,
\bq
\mrD = \eG{2 - \frac{1}{2}\,\ep}\,\frac{\pi^{\ep/2}}{1 - \ep/2}\,\Bigl( \mrD_1 + \mrD_2) \spc \quad
\mrD_\mrj = \int_0^1 \mrd \mrx_2 \, \int_0^{\mrx_2} \mrd \mrx_3\,
\mrQ_\mrj^{ - 1 + \ep/2}\,\mrL^{-1} \spc
\eq
\bqa
\mrQ_1 &=& 
\mrm^2\,(\mrx_2^2 + \mrx_3^2) + 
(\mrs - 2\,\mrm^2)\,\mrx_2\,\mrx_3 - 
2\,\mrm^2\,\mrx_2 +
(\mrt + \mru - \MHs)\,\mrx_3 + 
\mrm^2 \spc
\nl
\mrQ_2 &=& \mrt\,\mrx_2^2 + \mrm^2\,\mrx_3^2 +
(\mrs + \mru - 3\,\mrm^2)\,\mrx_2\,\mrx_3 -
(\mrt + \mrm^2)\,\mrx_2 +
(\mrt + \mrm^2 - \MHs)\,\mrx_3 + \mrm^2 \spc
\nl
\mrL &=& (\mrt - \mrm^2)\,\mrx_2 + (\mru - \mrm^2)\,\mrx_3 \spp
\eqa
Let us consider the $\mrD_1$ integral: after performing a change of variable,
\bq
\mrx_3 = \mrx^{\prime}_3 + \alpha\,\mrx_2 \spc \qquad
\alpha = - \frac{1 - \lambda}{1 + \lambda} \spc
\eq
where $\alpha \in \Rf$, we can perform the $\mrx_2$ integration giving a result that can be written
in terms of $\mrF^{(2)}_{\mrD}$ functions.
\bqa
\mcD_1 &=& \int_0^1\, \mrd \mrx_3\,\Bigl[
- \mrF^{(2)}_{\mrD}(1\,;\,1,1 - 1/2\,\ep\,;\,2\,;\,\frac{\mrd_1}{\mrd_1 + \mrc_2},\frac{\mrb_2}{\mrb_2 + \mra_2})\,
          (\mrd_1 + \mrc_2)^{-1}\,(\mrb_2 + \mra_2)^{-1+\ep/2}\,\alpha
\nl          
{}&-& \mrF^{(2)}_{\mrD}(1\,;\,1,1 - 1/2\,\ep\,;\,2\,;\, - \frac{\mrd_1}{\mrc_1}, - \frac{\mrb_1}{\mra_1})\,
          \mrc_1^{-1}\,\mra_1^{-1+\ep/2}\,(1 - \alpha)
\nl          
{}&+& \mrF^{(2)}_{\mrD}(1\,;\,1,1 - 1/2\,\ep\,;\,2\,;\, - \frac{\mrd_1}{\mrc1}\,\mrx_3, - \frac{\mrb_1}{\mra_1}\,\mrx_3) \, 
          \mrc_1^{-1}\,\mrx_3\,\mra_1^{-1+\ep/2}\,(1 - \alpha)
\nl          
{}&+& \mrF^{(2)}_{\mrD}(1\,;\,1,1 - 1/2\,\ep\,;\,2\,;\, - \frac{\mrd_1}{\mrc_2}\,\mrx_3, - \frac{\mrb_2}{\mra_2}\,\mrx_3) \, 
          \mrc_2^{-1}\,\mrx_3\,\mra_2^{-1+\ep/2}\,\alpha \Bigr] \spp
\eqa
We have introduced the following quantities:
\bq
\mrc_1 = (1 - \alpha)\,(\mru - \mrm^2)\,\mrx_3 \spc \qquad
\mrc_2 = - \alpha\,(\mru - \mrm^2)\,\mrx_3 \spp
\eq
From the boundaries of the phase space we derive that $\alpha < 0$ and $\mru > \mrm^2$; therefore $\mrc_{1,2} > 0$.
We have also introduced
\bqa
\mrd_1 &=& \MHs - \mrs + 2\,\frac{\mrm^2 - \mru}{1 + \lambda} > 0 \spc
\nl
\mra_1 &=& \mrs\,\Bigl(\frac{1 - \lambda}{1 + \lambda}\Bigr)^2\,(\mrx_3 - \mry_1)\,(\mrx_3 - \mry_2) \spc \quad
\mrb_1 = 2\,\mrs\,\frac{\lambda}{1 + \lambda}\,(\mrx_3 - \mry_1) \spc
\nl
\mra_2 &=& \frac{1}{4}\,\mrs\,\frac{1 - \lambda)^3}{1 + \lambda}\,(\mrx_3 - 1)\,(\mrx_3 - \mry_3) \spc
\eqa
\bq
\mry_1 = \frac{1}{2}\,(1 - \lambda) \spc \quad
\mry_2 = \frac{1}{2}\,\frac{(1 + \lambda)^2}{1 - \lambda} \spc \quad
\mry_3 = \Bigl(\frac{1 + \lambda}{1 - \lambda}\Bigr)^2 \spp
\eq
We now consider the arguments of the Lauricella functions; it follows that
\bq
 - \frac{\mrd_1}{\mrc_2} < 0 \spc \qquad - \frac{\mrd_1}{\mrc_2} < 0 \spp
\eq
The next argument is 
\bq
\zeta_1 = - \frac{\mrb_1}{\mra_1} = 2\,\lambda\,\frac{1 + \lambda}{(1 - \lambda)^2}\,\frac{1}{\mry_2 - \mrx_3} \spp
\eq
The maximum value of $\lambda^2$ is $1 - 4\,\mrm^2/\MHs$; introducing $\rho= 4\,\mrm^2/\MHs \muchless 1$ we have
$0 < \lambda < 1 - \rho/2$, \ie $\frac{1}{2} \le \mry_2 \le 4/\rho$. When we are close to the threshold $\mrs = 4\,\mrm^2$
we have
\bq
\mrx_3 < \mry_2 \;\to\; \zeta_1 > 0 \spc \qquad \mrx_3 > \mry_2 \;\to\; \zeta_1  \spp
\eq
Furthermore we easily derive that for $\mrx_3 < \mry_2$ we have $\mrz_1 > 1$. It follows that there are two scenarios: 
with $\mrz_1 > 1$ where we do not transform the Lauricella functions; with $\mrz_1 < 1$ we need to transform the
Lauricella functions before introducing the corresponding MB representation. The last argument is
\bq
\zeta_2 = - \frac{\mrb_2}{\mra_2} = 4\,\frac{\lambda}{(1 - \lambda)^3}\,\frac{1}{\mry_3 - \mrx_3} \spp
\eq
We have $\mry_3 > 1$, therefore there are values of $\lambda$ where $0 < \zeta_2 < 1$; in those cases we need to
transform before introducing the MB representation. The remaining integral over $\mrx_3$ introduces again
Lauricella functions where transformations will be applied (whe needed) to derive the final MB representation.
\subsection{Strategy \label{stra}}
The examples discussed in this Section illustrate the general strategy.
Consider a quadratic form in $\mrN$ variables,
\bq
\mrV = (\mrx - \mrX)^{\mrt}\,\mrH\,(\mrx - \mrX) + \Delta \spp
\eq
It follows that $\Delta = 0$ induces a pinch on the integration contour
\bq
 \mrI = \int_0^1\,\mrd \mrx_1\,\dots\,
 \int_0^{\mrx_{\mrN-1}}\,\mrd \mrx_{\mrN}\,\mrV^{\alpha} \spc
\eq
at $\mathbf{x} = \mathbf{X}$ if
$0 < \mrX_{\mrN} < \,\dots\, < \mrX_1 < 1$
We obtain the leading singularity, usually called the anomalous threshold
(herefater AT). Sub{-}leading singularities are the leading singularities
of diagrams obtained by shrinking to a point one (or more) lines of the
original diagram. Consider the general one{-}loop triangle with
propagators (and Feynman parameters)
\[
\begin{array}{ccc}
\mrq_1^2 + \mrm_1^2 \quad & \quad (\mrq + \mrp_1)^2 + \mrm_2^2\quad & \quad
(\mrq + \mrp_1 + \mrp_2)^2 + \mrm_3^2 \\
\mry \quad & \quad \mrx - \mry \quad & \quad 1 - \mrx \\
\end{array}
\]
with $\mrp_{\mrj}^2 = - \mrM_{\mrj}^2$ and $(\mrp_1 + \mrp_2)^2 = - \mrs$.
In almost all cases the anomalous threshold lies outside the physical region. The other
singularities can be described as follows:

\begin{enumerate}

\item we shrink the line with propagator $(\mrq + \mrp_1)^2 +
\mrm_2^2$ to a point, corresponding to $\mry = \mrx$. Therefore, we
change variables, $\mrx^{\prime} = \mrx$, $\mry^{\prime} = \mrx - \mry$, obtaining
\bq
\mrC = \int_0^1 \mrd \mrx\,\int_0^{\mrx} \mrd \mry\,
\Bigl[ \mry\,\mrL(\mrx\,,\,\mry) + \mrQ(\mrx) \Bigr]^{-1} \spc
\eq
\bq
\mrL = (\mrM_2^2 - \mrM_1^2 - \mrs)\,\mrx + \mrM_1^2\,\mry + \mrs -
\mrM_2^2 +\mrm_2^2 - \mrm_1^2 \spc
\eq
\bq
 \mrQ = \mrs\,\mrx^2 + (\mrm_1^2 - \mrm_3^2 - \mrs )\,\mrx + \mrm_3^2 =
        \mrs\,(\mrx - \mrX)^2 - \frac{1}{4\,\mrs} \,
 \lambda\bigl(\mrs\,,\,\mrm_1^2\,,\,\mrm_3^2 \bigr) \spc
\eq
where $\mrX = (\mrs + \mrm_3^2 - \mrm_1^2)/(2\,\mrs)$. Therefore
$\lambda = 0$ gives a zero or $\mry = 0$ ad $\mrx = \mrX$. The singularity
appears for $\mrs = (\mrm_1 \pm \mrm_3)^2$ if $0 < \mrX < 1$, i.e. the
normal (pseudo) threshold. How to proceed in order to derive a MB
representation depends on the signs of $\lambda$ and of $\mrL$.

\item We can also shrink to a point the line corresponding  to the
propagator $\mrq^2 +\mrm_1^2$, corresponding to $\mry = 0$. In this case
\bq
\mrQ = \mrM_2^2\,(\mrx - \mrX)^2 -
\frac{1}{4\,\mrM_2^2}\,\lambda(\mrM_2^2\,,\,\mrm_2^2\,,\,\mrm_3^2)
\quad \mrX = - \frac{1}{2\,\mrm_2^2}\,(\mrm_2^2 - \mrm_3^2 - \mrM_2^2) \spc
\eq
and we can discuss another branch of the integral corresponding to
$\mrM_2^2 = (\mrm_2 \pm \mrm_3)^2$.

\end{enumerate}

A more complicated example is the kite integral, described by propagators
\[
\begin{array}{lll}
\mrq_1^2 + \mrm_1^2 \quad & \quad
(\mrq_1 + \mrp)^2 + \mrm_2^2 \quad & \quad
(\mrq_1 - \mrq_2)^2 + \mrm_3^2 \\
\mrq_2^2 + \mrm_4^3 \quad & \quad
(\mrq_2 + \mrp)^2 + \mrm_5^2 & \\
\end{array}
\]
The dimensionless second Symanzik polynomial for $\mrm_\mrj = \mrm$ is
\bq
\mrQ + \beta\,\lpar \lambda^2 - \frac{1}{4}\,\sigma_3 \rpar \spc \quad
\mrQ = \rho_1^2\,\sigma_1\,\lpar \mrx_1 - \frac{1}{2} \rpar^2 + 
       \rho_2^2\,\sigma_2\,\lpar \mrx_2 - \frac{1}{2} \rpar^2 +
       2\,\rho_1\,\rho_2\,\rho_3\,\lpar \mrx_1 - \frac{1}{2} \rpar\,\lpar \mrx_2 - \frac{1}{2} \rpar \spp
\eq
The three{-}particle cut of the kite integral can be studied by observing
that the (equal mass) sunrise has the same cut and corresponds to a
contraction of the kite with $\mrx_1 = 1$ and $\mrx_2 = 0$. It is
convinient to introduce new variables,
\bq
\mru_1 = 1 - \mrx_1 \spc \quad
\mru_2 = \mrx_2 \spc \quad
\mru_3 = \rho_1 - \frac{1}{3} \spc \quad
\mru_4= \rho_2 - \frac{1}{3} \spc
\eq
and rewrite $\mrS_2$ as
\bq
\frac{1}{\mrs}\,\mrS_2 = \mrQ(\mru_1\,,\,\mru_2) + \frac{1}{3}\,(\lambda - \frac{1}{9}) \spc
\eq
\bq
\mrQ = - \mrr_1\,\mru_1^2 - \mrr_2\,\mru_2^2
- 2\,\lpar \mrr_3 + \frac{1}{27} \rpar\,\mru_1\,\mru_2
+ \lpar \mrr_1 + \mrr_3 + \frac{1}{27} \rpar\,\mru_1
+ \lpar \mrr_2 + \mrr_3 + \frac{1}{27} \rpar\,\mru_2
- \lambda\,\mrr_4 - \mrr_3 \spc
\eq
\bqa
\mrr_1 &=& \mru_3^3 - \frac{1}{3}\,\mru_3 - \frac{2}{27} \spc
\qquad
\mrr_2 = \mru_4^3 - \frac{1}{3}\,\mru_4 - \frac{2}{27} \spc
\nl
\mrr_3 &=& - \mru_3\,\mru_4^2 - \mru_3^2\,\mru_4 - \frac{1}{3}\,\mru_4 \spc
\qquad
\mrr_4 = \mru_3^2 + \mru_4^2 + \mru_3\,\mru_4 \spp
\eqa
Therefore the point $\mru_\mrj = 0$ is a zero of the Symanzik polynomial
for $\lambda = 1/9$, i.e. the normal (pseudo) threshold of the kite
integral. We can now proceed below and above the normal thrshold with a
procedure depending on the sign of $\mrQ$.

To summarize: the procedure is as follows:
given $\mrN$ propagators introduce $\mrN$ parameters $\alpha_\mrj$ with
$\sum_{\mrj}\,\alpha_\mrj = 1$ and
\begin{enumerate}

\item  select the sub{-}diagram where the leading singularity is the
relevant parameter,

\item set the relevant $\alpha_{\{\mrj\}}$ to zero,

\item find the solution of the Landau equations for the remaining
$\alpha_{\{\mrk\}}$ parameters and shift them, i.e. if $\alpha_{\mrk} =
\mrA_{\mrk}$, with $\mrA_{\mrk}$
solution of the corresponding set of reduced Landau equations, shift
$\alpha_{\mrk} = \mrA_{\mrk} + \beta_{\mrk}$. At this point we can use the
partial quadratization of the Symanzik polynomials introducing
$\mathbf{x}$ and $\mathbf{\rho}$ variables instead of the $\mathbf{\beta}$
variables

\item Use MB splitting and MB representations only whe they are allowed;
when needed use relations among generalized hypergeometric functions
before introducing MB representions.

\end{enumerate}

Steps $1{-}4$  have been described in details in Sect.~$7$ of \Bref{Passarino:2024ugq}.
\section{Conclusions \label{conc}}
In this work we have studied the problem of writing the Mellin{-}Barnes representation (akas Fox functions)
of Feynman integrals corresponding to physical processes and taking into account their behavior below and
above the normal thresholds characterizing the integrals. 
We have described the use of contiguity relations and of Kummer transformations for generalized,
multivariate hypergeometric functions.
The numerical computation of Fox functions will be the argument of a forthcoming paper~\cite{NCFF}. 
\clearpage
\bibliographystyle{elsarticle-num}
\bibliography{FFPR}
\end{document}